\documentclass[]{aa} 

\begin{document}

\title{Phase mixing of nonlinear Alfv\'en waves}
\author{A.~P.~K. Prokopyszyn \inst{1} \and A.~W. Hood \inst{1} \and I. De Moortel \inst{1}}

\institute{School of Mathematics and Statistics, University of St Andrews, St Andrews, Fife, KY16 9SS, U.K.
\\
\email{apkp@st-andrews.ac.uk}}

\abstract
{}
{This paper presents 2.5D numerical experiments of Alfv\'en wave phase mixing and aims to assess the effects of nonlinearities on wave behaviour and dissipation. In addition, this paper aims to quantify how effective the model
presented in this work is at providing energy to the coronal volume.}
{The model is presented and explored through the use of several numerical experiments which were carried out using the Lare2D code. The experiments study footpoint driven Alfv\'en waves in the neighbourhood of a two-dimensional x-type null point with initially uniform density and plasma pressure. A continuous sinusoidal driver with a constant frequency is used. Each experiment uses different driver amplitudes to compare weakly nonlinear experiments with linear experiments.}
{We find that the wave trains phase-mix owing to variations in the length of each field line and variations in the field strength. The nonlinearities reduce the amount of energy entering the domain, as they reduce the effectiveness of the driver, but they have relatively little effect on the damping rate (for the range of amplitudes studied). The nonlinearities produce density structures which change the natural frequencies of the field lines and hence cause the resonant locations to move. The shifting of the resonant location causes the Poynting flux associated with the driver to decrease. Reducing the magnetic diffusivity increases the energy build-up on the resonant field lines, however, it has little effect on the total amount of energy entering the system. From an order of magnitude estimate, we show that the Poynting flux in our experiments is comparable to the energy requirements of the quiet Sun corona. However a (possibly unphysically) large amount of magnetic diffusion was used however and it remains unclear if the model is able to provide enough energy under actual coronal conditions.}
{}



\maketitle

\section{Introduction}

It has been hypothesised as early as \citet{Schatzman1949} that the solar corona could be heated largely by the dissipation of magnetohydrodynamic (MHD) waves generated in the lower layers of the Sun. Heating by MHD waves is still one of the mechanisms under consideration for heating the corona; see for example \citet{Klimchuk2006}, \citet{Parnell2012}, \citet{DeMoortel2015} and \citet{Klimchuk2015} for an overview of the coronal heating problem and the open questions that still need to be addressed. Magnetohydrodynamic waves are commonplace in the solar atmosphere and have been observed over the last two decades as a consequence of improved imaging and spectroscopic instruments (see e.g. \citet{DeMoortel2012}). 

A review of the linear behaviour of MHD waves can be found in, for example \citet{Goossens2011}. The dissipation of Alfv\'en waves has been the basis of many coronal heating models (see review by \citet{Arregui2008}, \citet{Arregui2015} and references therein). The main mechanisms for converting the energy associated with Alfv\'en waves into heat are Ohmic and viscous dissipation. Both of these heating mechanisms are proportional to gradients in either magnetic field or velocity. The steeper the gradients, the more efficient the wave energy is converted into thermal energy. Two main mechanisms for generating steep gradients have been proposed, namely, phase mixing \citep{Heyvaerts1983} and resonant absorption \citep{Ionson1978}. Phase mixing occurs when Alfv\'en waves propagating on magnetic surfaces move out of phase with neighbouring waves on nearby surfaces and this means that steep cross field gradients form. Resonant absorption occurs when driven standing Alfv\'en waves resonate while their neighbours on different magnetic surfaces are not resonating. Other mechanisms such as a turbulent cascade of wave energy (e.g. \citet{Hollweg1986}, \citet{Cranmer2007} and \citet{VanBallegooijen2011}) or coupling with compressive wave modes (\citet{Kudoh1999} and \citet{Antolin2010}) have also been proposed to increase the damping rate of Alfv\'en waves.

Phase mixing has been investigated as a mechanism for heating in many parts of the atmosphere, including coronal holes \citep[e.g.][]{Hood2002} and inside flux tubes \citep[e.g.][]{Pagano2018}. In this paper, we evaluate the effect of nonlinearities on the amount of energy which can be provided to a coronal domain by Alfv\'en waves to account for the energy lost through optically thin radiation and thermal conduction. The energy required to keep a loop at coronal temperatures has been researched extensively (see for example \citet{Rosner1978}, \citet{Martens2010}, \citet{Priest2014} and references therein). 

Nonlinear effects have been studied in a variety of settings. The magnetic tension force associated with an Alfv\'en wave is a linear force whereas the associated magnetic pressure force is a nonlinear force, often called the ponderomotive force (e.g. \citet{Verwichte1999}.
 
In this paper, our model combines several nonlinear effects, the most important of which is the generation of density structures (e.g. \citet{Terradas2004}).
Other nonlinear effects considered in our model are as follows: nonlinear coupling from Alfv\'en waves to magnetoacoustic waves (e.g. \citet{Verwichte1999}, \citet{Tsiklauri2001} and \citet{Thurgood2013ponderomotive} and Alfv\'en wave steepening due to the Alfv\'en speed being dependent on the perturbed magnetic energy associated with the Alfv\'en wave (e.g. \citet{Verwichte1999} and \citet{Tsiklauri2002}).
However, these latter effects appear to play a less significant role.

The generation of density structures and the coupling to magnetoacoustic waves are generated by the ponderomotive force. \citet{Verwichte1999}
showed that every time two Alfv\'en pulses superimpose each other they generate slow magnetoacoustic waves (in $\beta\ll1$ plasma) due to a force they call the cross-ponderomotive force, which is a subset of the nonlinear magnetic pressure force. The same is true whenever Alfv\'en waves reflect off a solid boundary. \citet{Thurgood2013ponderomotive} showed that if there are gradients in the Alfv\'en speed perpendicular to the magnetic field then fast waves are generated. \citet{Terradas2004} showed that a line-tied standing Alfv\'en wave pushes plasma towards the antinodes and away from the nodes due to the nonlinear magnetic pressure force and this creates a loop aligned density profile. They showed that in a $\beta=0$ plasma, the amplitude of the generated density profile grows algebraically with a $t^2$ profile, but this growth is limited by the plasma pressure force if $\beta\ne0$.
 
In this paper, we focus on the nonlinear aspects of Alfv\'en wave propagation and dissipation in a simplified version of a coronal arcade system. The paper is structured as follows: In Section \ref{sec:method} the phase mixing model is presented and a linear and ideal solution to the model is calculated, which is used to compare with the nonlinear experiments. In Section \ref{sec:results} the linear results are presented. Section \ref{sec:nonlinearities} assesses how the nonlinearities affect the heating in the model. Section \ref{sec:discussion} is a discussion on quantifying how effective the system is at converting wave energy into heat for typical coronal values. Finally, in Section \ref{sec:conclusions} conclusions are presented.

\section{Method and set-up}
\label{sec:method}
\subsection{Equations}
\label{sec:equations}

All the numerical experiments presented in this paper are performed using the MHD code Lare2D \citep{Arber2001}. The code solves the following set of MHD equations:
\begin{equation}
\label{eq:continuity}
    \frac{\text{D}\rho}{\text{D}t} = -\rho \vec{\nabla} \cdot \vec{v}, 
\end{equation}
\begin{equation} 
\label{eq:momentum} 
    \rho \frac{{\text{D}\vec{v}}}{{\text{D}t}} = \vec{j} \times \vec{B} - \vec{\nabla} p + \vec{F}_\nu^{shock}, \end{equation}
\begin{equation}
\label{eq:energy}
    \rho \frac{{\text{D}\epsilon}}{{\text{D}t}} = - p(\vec{\nabla} \cdot \vec{v})+ j^2/\sigma + H_\nu^{shock}, \end{equation}
\begin{equation}
\label{eq:iduction}
\frac{\text{D}\vec{B}}{\text{D}t}=\left(\vec{B} \cdot \vec{\nabla}\right)\vec{v} - \left(\vec{\nabla} \cdot \vec{v} \right) \vec{B} + \eta\nabla^2\vec{B}, 
\end{equation}
\begin{equation}
    p = \frac{k_B}{\mu_m m_p}\rho T
,\end{equation}
where $\vec{F}_\nu^{shock}$ and $H_\nu^{shock}$ are terms related to the shock viscosity of the code, which is based on the edge viscosity formulation in \citet{Caramana1998}. The Boltzmann constant is denoted with $k_B$, the mass of a proton is denoted with $m_p$, and the mass fraction of the ions in proton masses is denoted with $\mu_m=1/2$. All other variables have their usual meanings. The code uses a uniform value for the magnetic diffusivity, $\eta$, given by
\begin{equation}
    \label{eq:eta}
    \eta = 10^{-3}\eta_0,
\end{equation}
where
$$\eta_0 = \frac{L_0B_0}{\sqrt{\mu\rho_0}},$$
where $B_0$, $L_0$, and $\rho_0$ are normalising constants and where $\rho_0$ also corresponds to the initial density. The value used for $\eta$ is unphysically large, but was chosen to be as small as possible without the effects of numerical diffusion and dispersion becoming too large. In the corona, the value of $\eta$ is roughly equal to 1$\,\text{m}^2\,s^{-1}$ \citep[see][p.~79]{Priest2014}, which means that for $\eta$ in the code to be physically accurate, it should be approximately
$$\eta=10^{-12}\eta_0,$$
if $L_0=1$Mm, $B_0=10^{-3}$T, and $\rho_0=10^{-12}$kg\,$\text{m}^{-3}$. The effects of varying the parameter $\eta$ are explored in Section \ref{sec:discussion}. In our experiments, the coefficient of kinematic viscosity is set to zero. Hence, there is no heating due to this term. However, shock viscosities are included, along with any heating associated with shocks. This was chosen partly because according to \citet{VanDoorsselaere2007} observational evidence favours a resistive (wave) heating mechanism for coronal loops over viscous dissipation. In Section \ref{sec:location_of_heating}, some of the consequences of setting the coefficent of kinematic viscosity to zero are discussed.
\subsection{Initial conditions}
\begin{figure}
    \centering
    \includegraphics[width=0.49\textwidth]{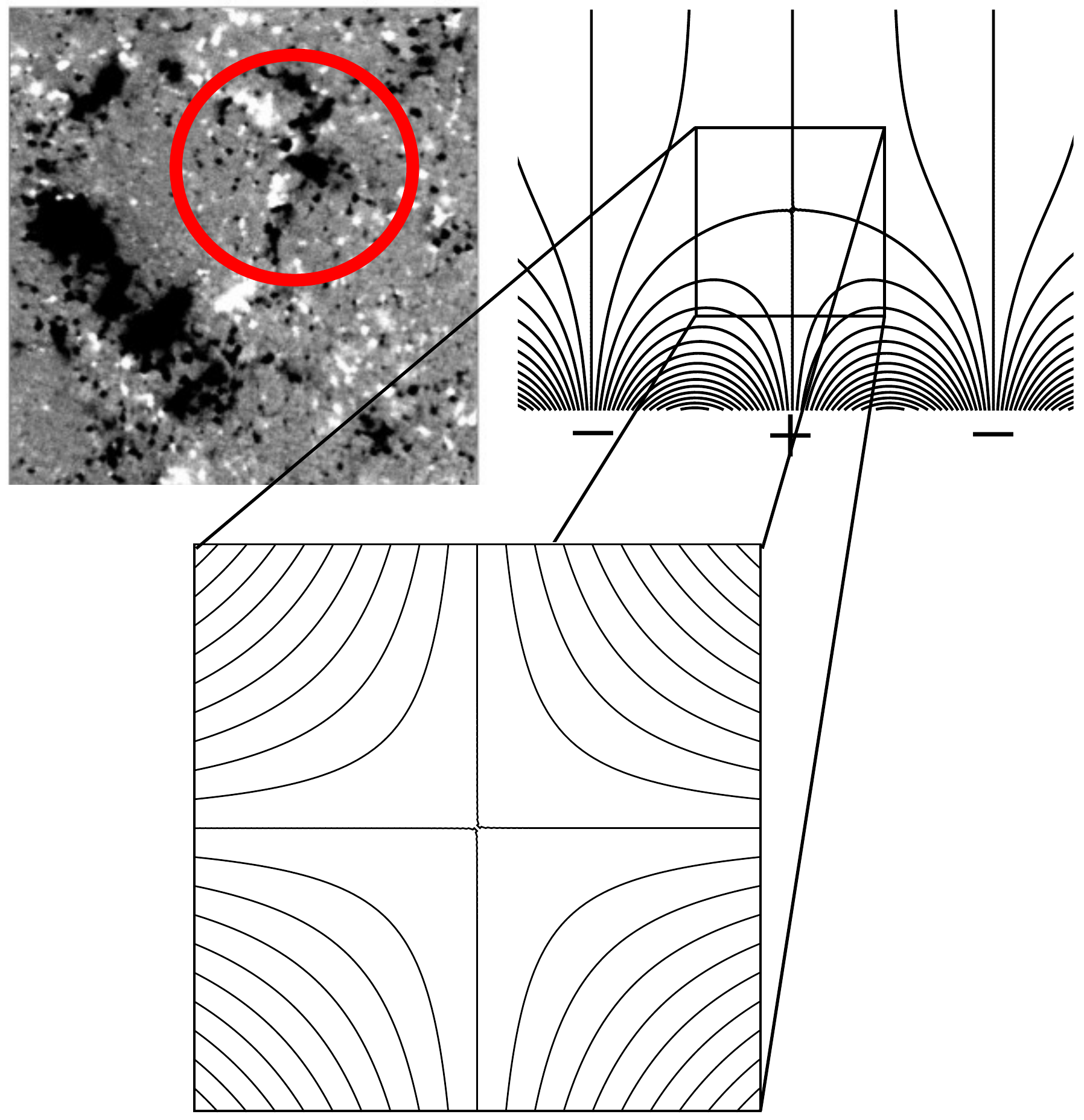}
    \caption{Top left figure: Magnetogram taken from the Hinode spacecraft of a mixed-polarity region. The top right figure shows a simplified diagram of the magnetic field configuration in a mixed polarity region when viewed edge on (for example when viewed on the solar limb). The bottom figure isolates the centre of the top right figure and is the profile of the magnetic field used in the numerical experiments of this paper.}
    \label{fig:magnetogram_exp_field_x_field}
\end{figure}

Phase mixing and resonant absorption require gradients in the Alfv\'en speed and this is often achieved through the use of an imposed density profile. A recent paper by \citet{Cargill2016} demonstrated that wave heating cannot sustain the assumed density structure. For this reason, we do not assume any density structures in this paper and instead, a uniform initial density profile is used. Our experiment relies on a gradient in magnetic field strength and also a variation in field line length to provide the conditions necessary for phase mixing. 

An initial static equilibrium is set up with uniform density ($\rho_0$) and pressure ($p_0$). The initial magnetic field is a potential 2D x-type null point, $\vec{B}_0$, defined as
\begin{equation}
\label{eq:magnetic_field}
    \vec{B}_0=\frac{B_0}{L_0}(x,-y,0).
\end{equation}
The $z$-direction is taken as the invariant direction, i.e. $\partial/\partial z \equiv 0$ throughout all the experiments. Our chosen magnetic field configuration is illustrated at the bottom of Figure \ref{fig:magnetogram_exp_field_x_field}. It was chosen to represent a simplified version of the magnetic field in a mixed polarity region. The top left image in Figure \ref{fig:magnetogram_exp_field_x_field} shows a magnetogram of a (generic) mixed polarity region and was taken from the Hinode spacecraft. A mixed polarity field was used because it contains strong variations in field strength and field line length. Two simplifications were made: the first is that the field is approximated using a 2D model and the second is that an x-type null point configuration was used. The x-point field is qualitatively similar to the top right field, particularly close to the null point. The top right field is a simplified diagram of the magnetic field in a mixed polarity region if viewed on the limb. Waves near an x-point field have been studied extensively; see for example \citet{McLaughlin2011}, \citet{McLaughlin2013}, \citet{Thurgood2013xpoint} and \citet{McLaughlin2016}. 

The initial uniform plasma pressure was chosen such that most of the domain is a low-beta domain. The $\beta=1$ contour is a circle occurring at a radius of $R/L_0=0.1$ about the null point. Since the magnetic field strength increases linearly with radius and the density is initially constant, the Alfv\'en speed increases linearly with radius.

\subsection{Boundary conditions}
\label{sec:boundary_conditions}

To simulate the steep jump in density and temperature between the chromosphere and the corona, reflective boundary conditions are used \citep[see][p.~434]{laney1998}. In other words, $\vec{v}=0$ and $\vec{\hat{n}}\cdot\nabla=0$ for all other variables, where $\vec{\hat{n}}$ is a vector normal to the boundary. 
The computational domain is given by
$$x_{min} \le x\le x_{max},\ y_{min} \le y \le y_{max},$$
where $x_{max}=y_{max}=2L_0$ and $x_{min}=y_{min}=-2L_0$. On the $y=y_{min}$ boundary, a continuous driver of the following form is imposed:
\begin{equation}
    \label{eq:driver}
    v_z = v_{driv}f(x)g(t),
\end{equation}
where $v_{driv}$ is the driver amplitude and $\tau_{driv}$ is the period of the driver. The spatial profile of the driver is described by the following equation:
\begin{equation}
    \label{eq:driver_spatial_profile}
    f(x)=
    \begin{cases}
        1 & |x|\le 1.5 L_0, \\
        \sin^2(\pi x/L_0) & 1.5L_0\le |x|\le 2.0 L_0.
    \end{cases}
\end{equation}
Equation \eqref{eq:driver_spatial_profile} implies that the spatial profile of the driver is constant along most of the boundary and smoothly goes to zero at the ends. To ensure the driver smoothly generates a wave train, the time profile of the driver is given by
\begin{equation}
    \label{eq:driver_time_profile}
    g(t) = 
    \begin{cases}
    \sin^2(2\pi t/\tau_{driv})& t \le \frac{1}{4}\tau_{driv}, \\
    \sin(2\pi t/\tau_{driv})&\frac{1}{4}\tau_{driv} \le t \le t_{end}^{driv}, \\
    0& t_{end}^{driv} \le t \le t_{end}. \\
    \end{cases}
\end{equation}
The period of the driver is given by
\begin{equation}
    \tag{\ref{eq:driving_time}}
    \tau_{driv}=\frac{L_0\sqrt{\mu\rho_0}}{B_0}4\log\left(2\right),
\end{equation}
where the function $\log(2)$ was chosen for convenience rather than any physical reason. In particular, it was chosen to be of the same form as the resonant field line locations (see Equation \eqref{eq:harmonic_time_periods}) such that the resonance occurs on field lines which are easier to describe analytically (see Section \ref{sec:resonance}).
After 20 driving time periods have elapsed, the driver is switched off and the experiments continue to run for another 5 driving time periods, such that
\begin{equation}
    t^{driv}_{end}=20\tau_{driv},
\end{equation}
\begin{equation}
    t_{end} = 25\tau_{driv}.
\end{equation}

One key simplification which is made is that the frequency spectrum of the driver is discrete, while the frequency spectrum of the driver which generates waves in the corona likely resembles that of a broadband spectrum. For reference see for example \citet{Wright1995}, \citet{DeGroof2002a}, \citet{DeGroof2002b}. In Section \ref{sec:discussion}, the effects of a more random driver are briefly discussed.

To enable us to assess how nonlinearities affect the energy evolution in the experiments, we analytically calculate the energy evolution in a similar set-up which is ideal and linear. The details of this calculation are given in Appendix \ref{sec:ideal_and_linear_solution_apdx}. When referring to energy values from the linear reference calculation, the following notation is used: $E_{lin}$ refers to the total energy input from the driver in the analytical ideal and linear set-up and $E_{lin}^{end}=E_{lin}(t_{end}^{driv})$ refers to the total energy input from the driver after the driving has finished. We note that $E_{lin}$ is a function of $v_{driv}$ and so each experiment is normalised by a different value depending on the value of $v_{driv}$. 

\section{Numerical results}
\label{sec:results}
Before discussing the nonlinear effects in Section \ref{sec:nonlinearities}, we describe the linear effects that occur in the experiments, which are obtained by using a velocity amplitude of $v_{driv}=10^{-3}v_{A0}$ in the numerical experiments. In order to reduce the number of Figures in the paper the graphs present linear and nonlinear results. The linear results are represented by solid blue lines and this section focusses on these results.

\subsection{Phase Mixing}

\begin{figure}
    \centering
    \includegraphics[width=.49\textwidth]{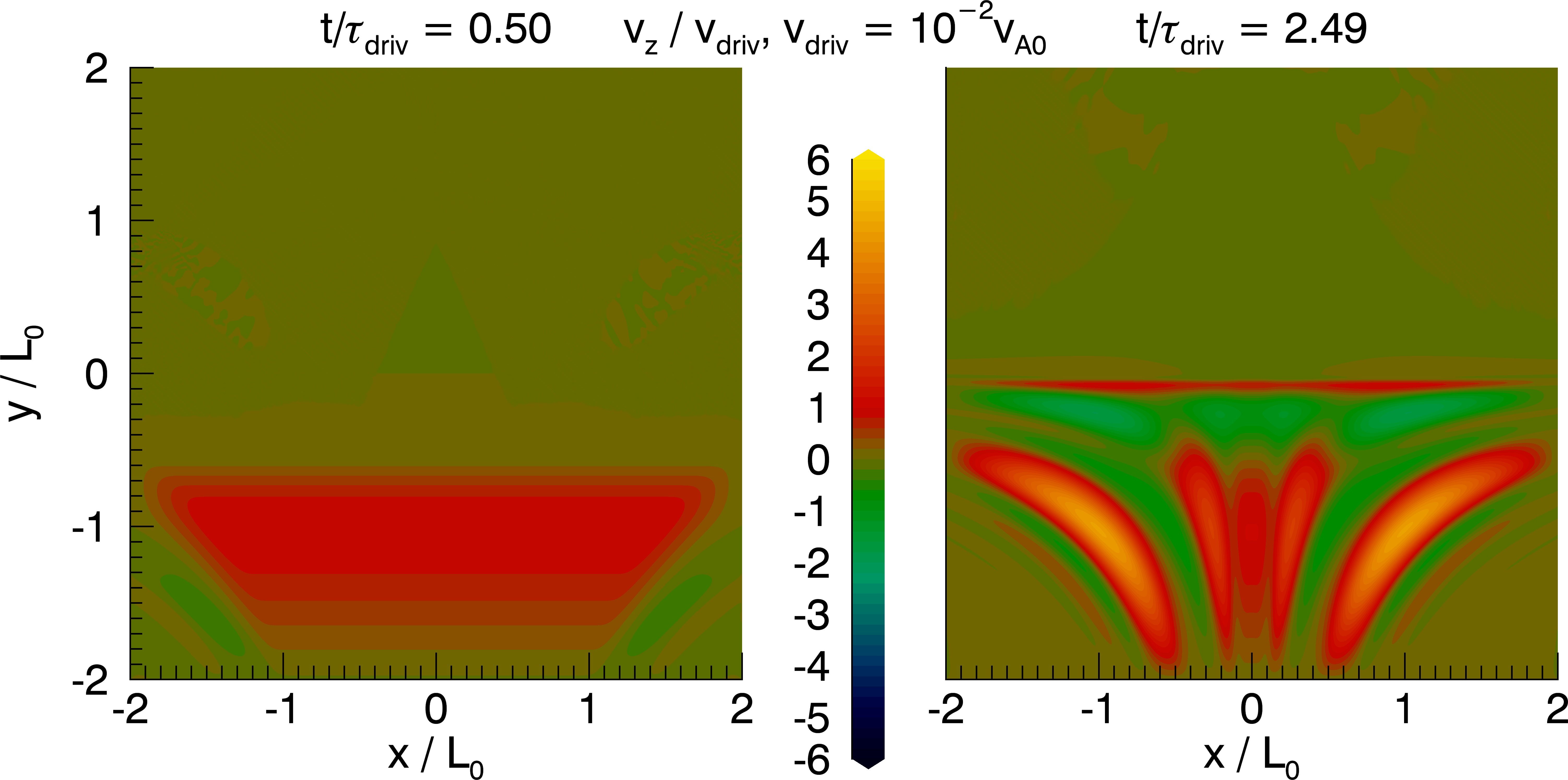}
    \caption{Contour plots of the (normalised) Alfv\'en wave velocity perturbations, $v_z$, at different times.}
    \label{fig:vz_contour_plots}
\end{figure}

\begin{figure}
    \centering
    \includegraphics[width=.49\textwidth]{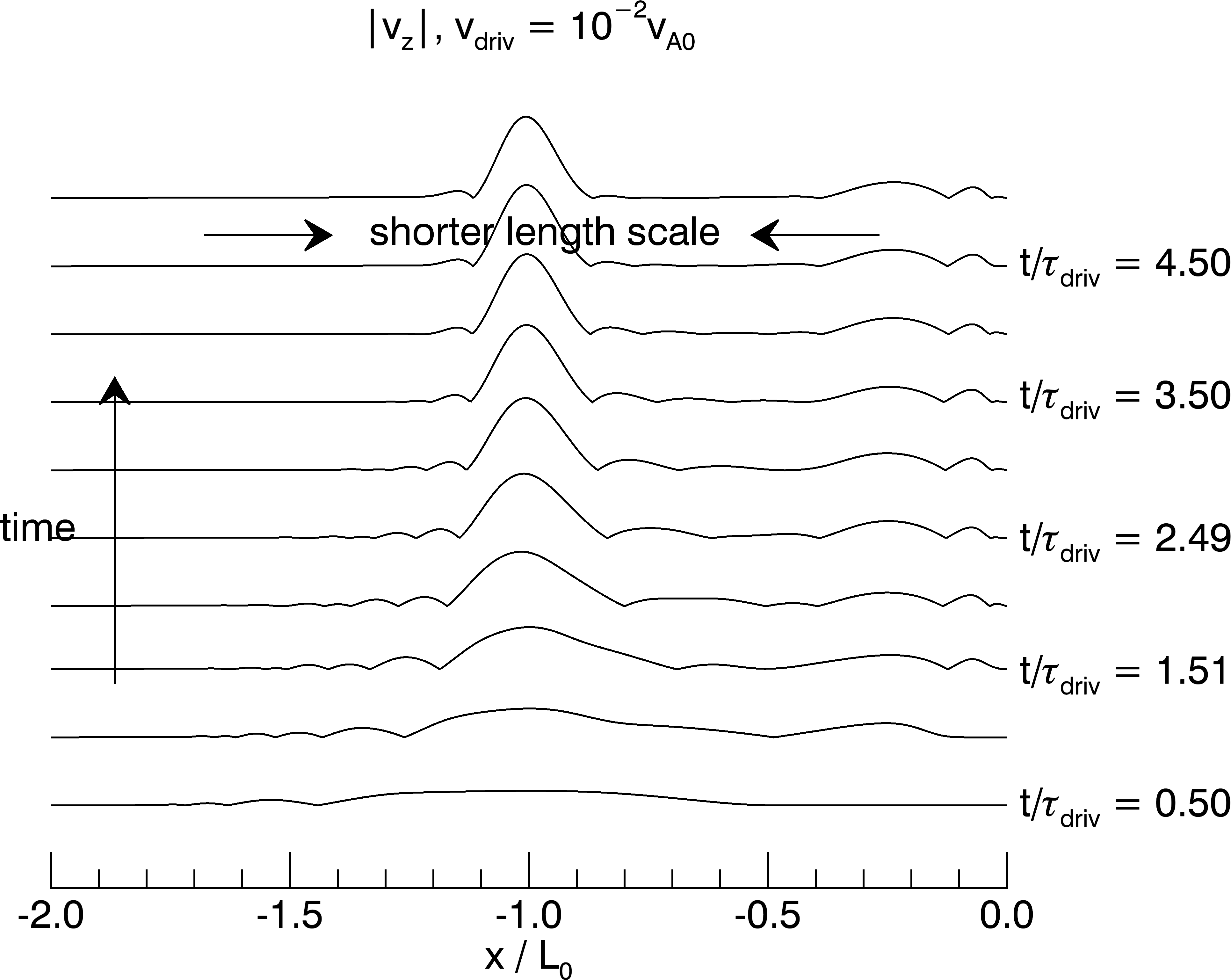}
    \caption{Absolute value of the velocity, $|v_z|$, associated with the Alfv\'en waves along the line $y=x$ at different times. The line $y=x$ is perpendicular to the field lines and this shows the variation in $v_z$ across the field lines. The figure shows that the length scale across the field lines is shortened as time progresses and so phase mixing is occurring.}
    \label{fig:vz_line_plots}
\end{figure}

\begin{figure}
    \centering
    \includegraphics[width=0.49\textwidth]{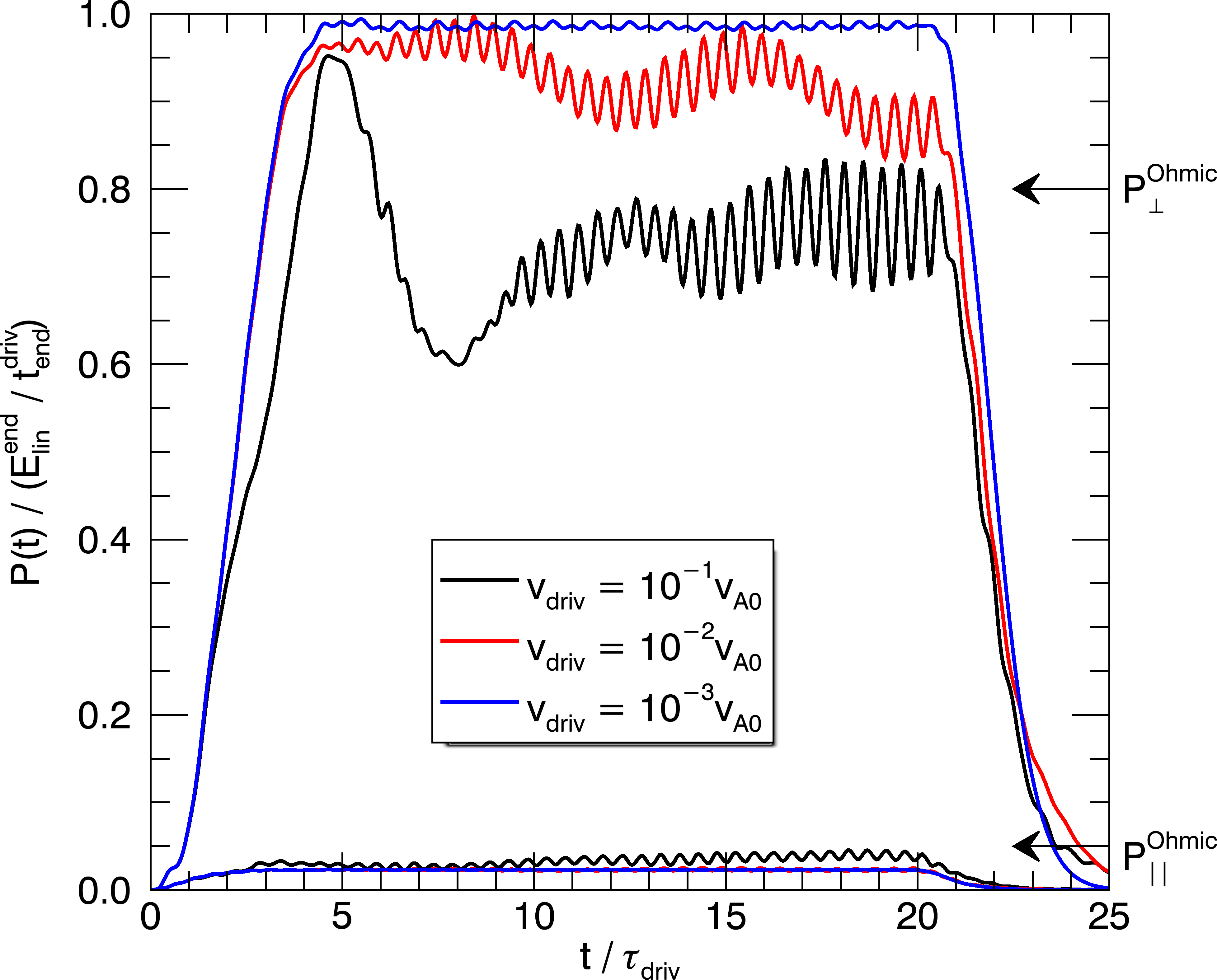}
    \caption{Plots of the Ohmic heating contributions from $\nabla_{||}B_z$ and $\nabla_{\perp}B_z$. Labels are provided on the right-hand side of the figure. The plots have been normalised by $E_{lin}^{end}/t_{end}^{driv}$ (see Section \ref{sec:boundary_conditions}), which gives the average power input from the driver in an equivalent but linear and ideal set-up.}
    \label{fig:perp_vs_par_ohmic_power}
\end{figure}

The left-hand side of Figure \ref{fig:vz_contour_plots} shows that the wave-front of the Alfv\'en wave is initially parallel to the $x$-axis. The right-hand side of the figure shows that at a later time, the Alfv\'en waves are out of phase with their neighbours. The phase mixing can also be seen in Figure \ref{fig:vz_line_plots}, which shows the velocity component of the Alfv\'en wave, $v_z$, along the line $y=x$ at multiple times. It can be seen that the length scale across the field lines has shortened owing to phase mixing. In this particular set-up, there are two reasons why the phase mixing occurs. The first reason is that there is a gradient in Alfv\'en speed across the field lines due to variations in magnetic field strength. The second reason is that there is a variation in the length of each field line and so different waves reflect at different times relative to their neighbours, again leading to out-of-phase waves on neighbouring field lines.

The remainder of this subsection shows that the dominant reason for the formation of magnetic field gradients is indeed because of phase mixing and not other effects such as wave steepening as the waves approach the null. Phase mixing generates gradients in the magnetic field across different magnetic field lines, whereas wave steepening generates gradients parallel to the magnetic field lines. In other words, phase mixing generates $\nabla_\perp B_z$ gradients whereas wave steepening generates $\nabla_{||}B_z$ gradients, namely,
$$\nabla_{||}=\frac{\vec{B}_0\cdot\nabla}{|\vec{B}_0|},\ \nabla_\perp = \frac{\vec{\hat{z}}\times\vec{B}_0\cdot\nabla}{|\vec{B}_0|},$$
where $||$ refers to the component parallel to the magnetic field and $\perp$ refers to the perpendicular component in the $\hat{z}\times\vec{B}_0$ direction. In Figure \ref{fig:perp_vs_par_ohmic_power}, the Ohmic heating contributions from both types of gradients are plotted, where
$$P_{||}^{Ohmic}=\frac{1}{\sigma}\left(\frac{\nabla_{||}B_z}{\mu}\right)^2,$$
$$P_{\perp}^{Ohmic}=\frac{1}{\sigma}\left(\frac{\nabla_{\perp}B_z}{\mu}\right)^2.$$
Figure \ref{fig:perp_vs_par_ohmic_power} clearly shows that substantially more heating occurs from gradients perpendicular to the magnetic field rather than parallel gradients, which confirms that phase mixing is indeed the dominant mechanism for generating the heating in this experiment.

\subsection{Resonance}
\label{sec:resonance}
\begin{figure}
    \centering
    \includegraphics[width=0.49\textwidth]{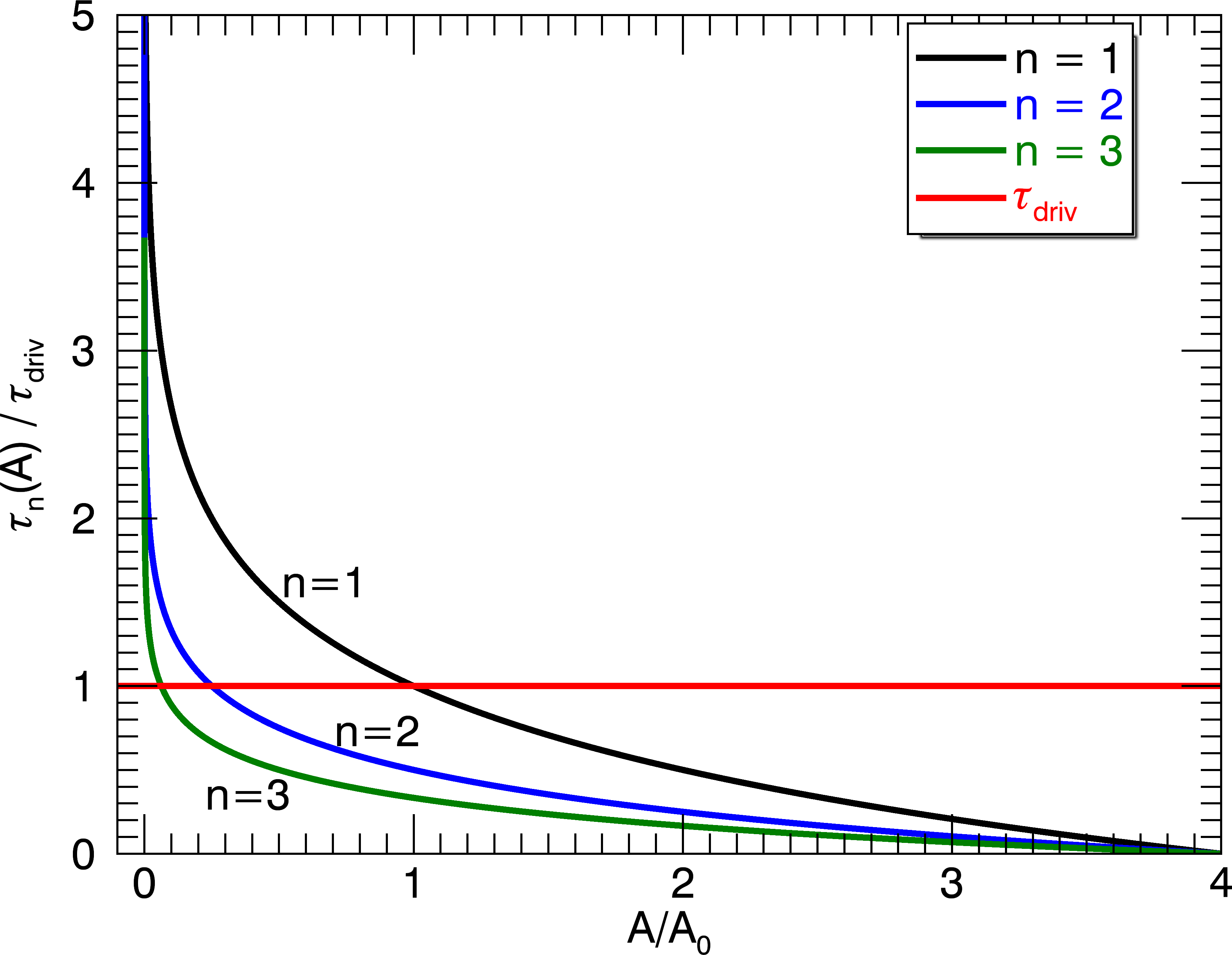}
    \caption{Plots of the first three harmonic periods ($\tau_n$) of each field as a function of the vector potential ($A$) normalised by the period of the driver ($\tau_{driv}$).}
    \label{fig:time_period}
\end{figure}
On the right-hand side of Figure \ref{fig:vz_contour_plots} and the final snapshots in Figure \ref{fig:vz_line_plots}, the signs of resonance occurring on discrete field lines can be seen. This section focusses on explaining why and where the resonance occurs. The harmonic time periods ($\tau_n$) of each field line at the initial time are given by the following equation:
\begin{equation}
    \tag{\ref{eq:harmonic_time_periods}}
    \tau_n = \frac{L_0\sqrt{\mu\rho_0}}{B_0}\frac{2}{n}\log\left(\frac{A_0}{L_0^2}\frac{x_{max}y_{max}}{|A|}\right),
\end{equation}
which is derived in Appendix \ref{sec:harmonic_time_periods}. In this case, $\vec{A}=A\vec{\hat{z}}$ is the vector potential of the magnetic field lines with flux function $A$, given by
\begin{equation}
    \tag{\ref{eq:vector_potential}}
    A=\frac{B_0}{L_0}xy\,.
\end{equation}
In 2D, lines of constant $A$ define the field lines and $A=0$ corresponds to the separatrices passing through the null point. The periods are plotted in Figure \ref{fig:time_period} as a function of $A$. The period of the driver, given by Equation \eqref{eq:driving_time}, is overplotted as the red horizontal line.
Resonance occurs on field lines where the period of the driver equals one of the harmonic periods. It can be seen that changing the period of the driver merely changes the location of the resonance but does not remove the resonance. In Appendix \ref{sec:resonance_locations}, the locations at which resonance occurs for these waves are shown to lie on field lines described by the following equation:
\begin{equation}
    \tag{\ref{eq:resonance_locations}}
    \frac{xy}{L_0^2}=\pm4^{1-n},\quad y\le0,
\end{equation}
where the integer $n$ is the harmonic number.
The above formula provides the resonance locations for linear waves in an ideal plasma. However, in Section \ref{sec:nonlinearities}, we show that some of the resonance locations shift outwards in the nonlinear experiments.

\subsection{Energy evolution and driver effectiveness}
\label{sec:poynting_flux}
\begin{figure}
    \centering
    \includegraphics[width=.49\textwidth]{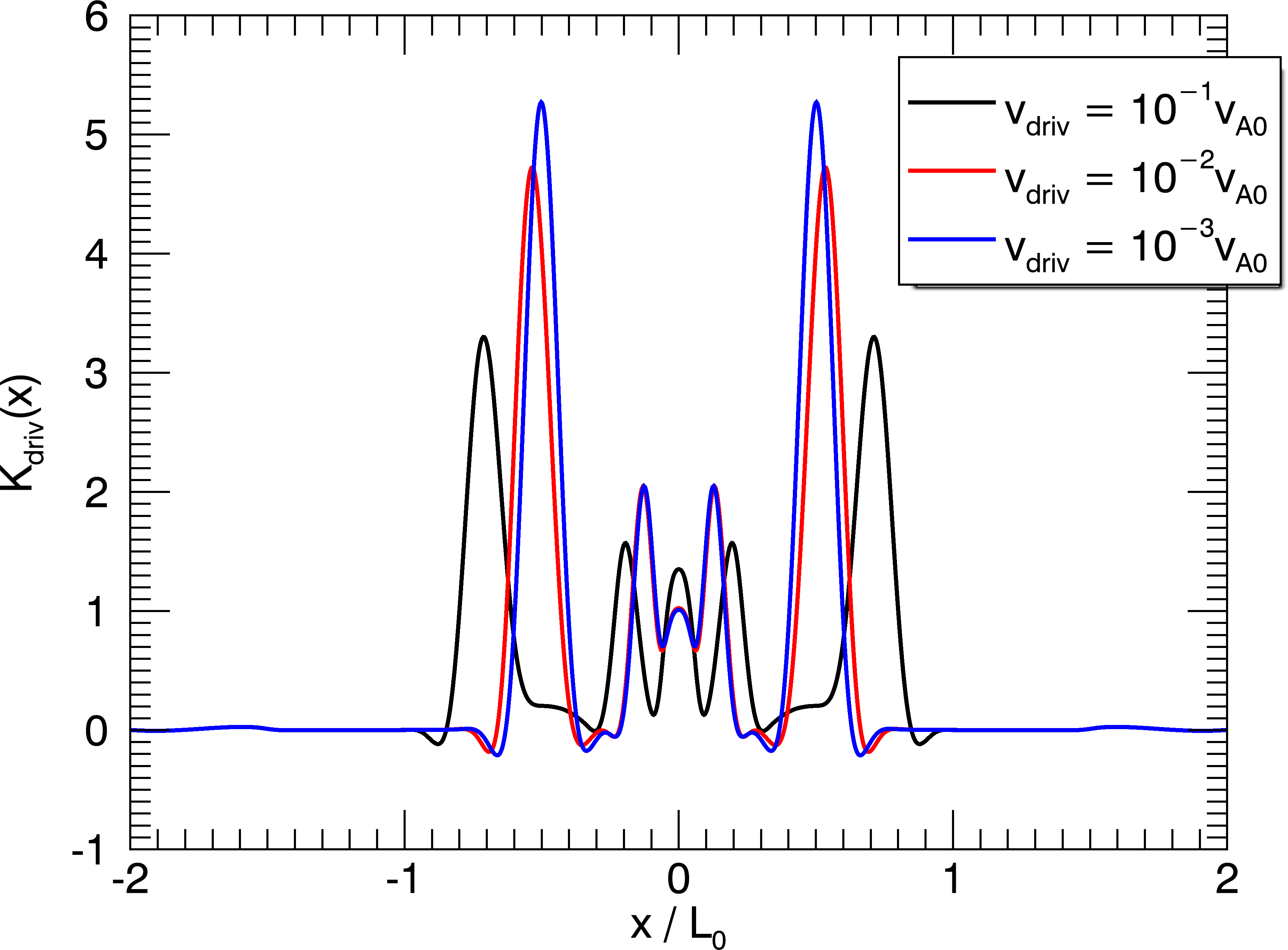}
    \caption{Plots of the driver effectiveness $K_{driv}$ (Eq.~\eqref{eq:driver_effectiveness}) on the bottom boundary.}
    \label{fig:poynting_flux}
\end{figure}
\begin{figure}
    \centering
     \includegraphics[width=0.49\textwidth]{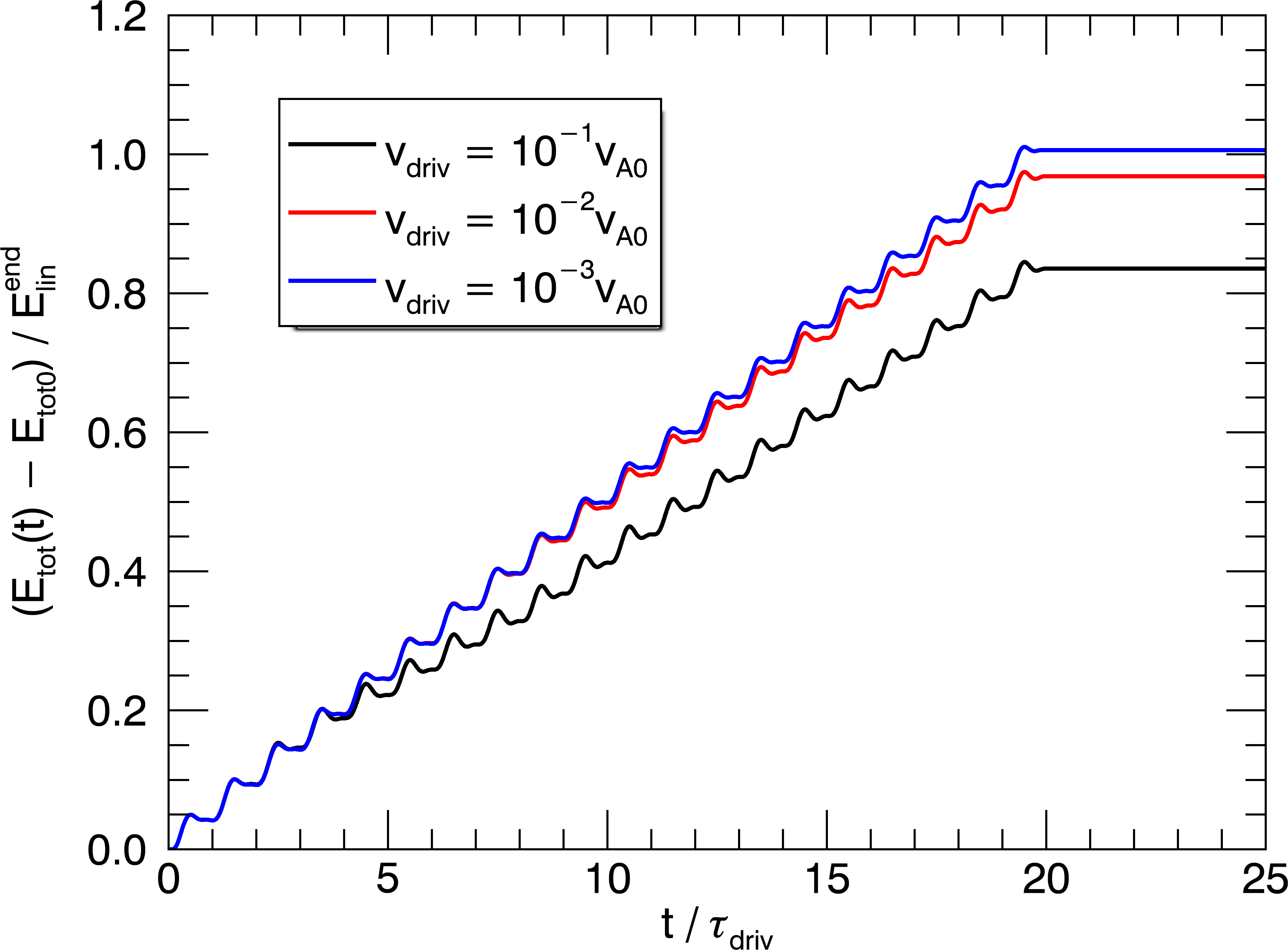}
    \caption{Plot of the change in total energy (Eq.~\eqref{eq:total_energy_equation0}) from its initial value ($E_{tot0}$) for different driver amplitudes. The plots have been normalised by $E_{lin}^{end}/t_{end}^{driv}$, which gives the total energy input from the driver in an equivalent but linear and ideal set-up.}
    \label{fig:total_energy}
\end{figure}
\begin{figure}
    \centering
    \includegraphics[width=.49\textwidth]{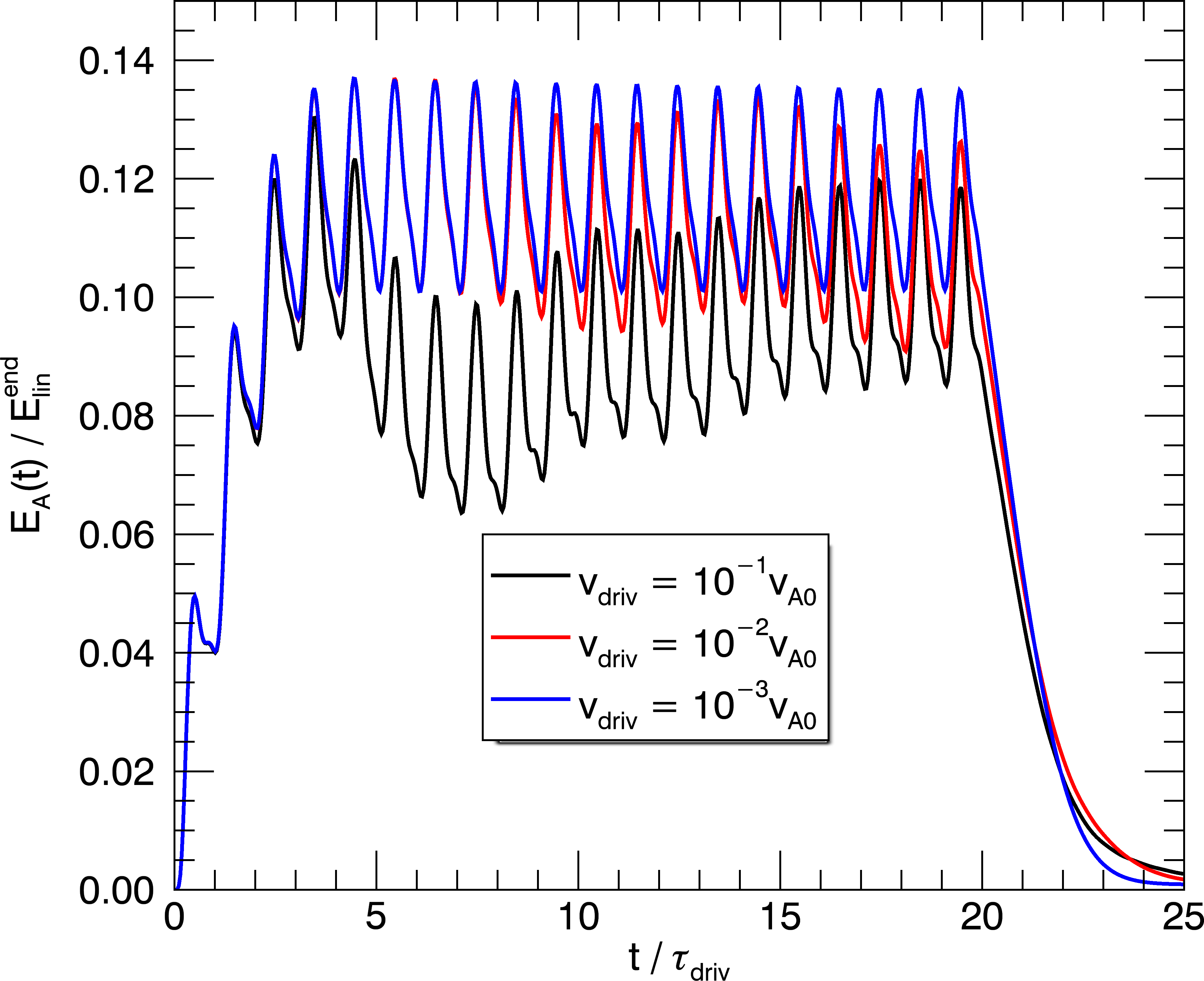}
    \caption{Plot of the Alfv\'en wave energy ($E_A$) for different driver amplitudes. The plots have been normalised by $E_{lin}^{end}/t_{end}^{driv}$, which gives the total energy input from the driver in an equivalent but linear and ideal experiment.}
    \label{fig:alfven_energy}
\end{figure}

With the exception of the driver, the velocity components are set to zero on all the boundaries.
This implies that the Poynting flux is the only term responsible for changes in the total energy of the system (see Appendix \ref{sec:total_energy_evolution} for proof). The change in total energy, $E_{tot}$, of the system is given by
\begin{equation}
    \tag{\ref{eq:total_energy_equation4}}
    \frac{dE_{tot}}{dt}=-\frac{B_y}{\mu}\int_{y=y_{min}}v_zB_zdx,
\end{equation}
where $E_{tot}$ is defined as
\begin{equation}
    \tag{\ref{eq:total_energy_equation0}}
    E_{tot}=\int_S\frac{p}{\gamma - 1}+\frac{B^2}{2\mu}+\frac{1}{2}\rho v^2dS,
\end{equation}
where $S$ is the computational domain.
The driver velocity, $v_z$, is imposed on the boundary by the driver given by Equation \eqref{eq:driver}. However, $B_z$ is free to adjust its value, which means the Poynting flux can be positive as well as negative. The value $B_y$ is positive and does not change on the driver boundary. Hence, the Poynting flux is determined by the relationship between $B_z$ and $v_z$. We introduce the dimensionless parameter $K(x,t)$, given by
$$\frac{B_z}{\sqrt{\mu}}=-K\sqrt{\rho_0}v_z,$$
which defines the relationship between $v_z$ and $B_z$. The equation $K=1$ corresponds to a single upward propagating linear Alfv\'en wave. We define the driver effectiveness, $K_{driv}(x)$, on the bottom boundary as
\begin{equation}
    \label{eq:driver_effectiveness}
    K_{driv}=\frac{1}{\sqrt{\rho_0\mu}}\frac{\int_0^{t_{end}^{driv}} B_zv_zdt}{\int_0^{t_{end}^{driv}} v_z^2dt},
\end{equation}
where $K_{driv}$ is a weighted time average of $K$. $K_{driv}$ gives a measure of how effectively the driver provides energy to a given field line. To see this more clearly, the above equation is equivalent to
$$K_{driv}=\frac{-B_y/\mu\int_0^{t_{end}^{driv}} v_zB_z dt}{-B_y/\mu\int_0^{t_{end}^{driv}} v_z(-\sqrt{\rho_0\mu}v_z)dt},$$
where the denominator corresponds to the Poynting flux through the boundary for  an upward propagating linear Alfv\'en wave.

For a field line which has only upward propagating linear Alfv\'en waves at the driven boundary, $B_z=-\sqrt{\mu\rho_0}v_z$, hence,  $K_{driv}=1$. A field line may only have upward propagating waves at the driven boundary because the waves are efficiently damped before they can reflect or the field is open. If the waves are reflected then the relationship is more complicated as there are now upward propagating and downward propagating waves at the driven boundary. In most cases, reflection acts to reduce driver effectiveness as reflection opens up the possibility for destructive interference to occur. However, if the frequency of the driver is such that it sets up a resonance, then this has the effect of increasing the driver effectiveness. This reflects the fact that resonant field lines have the property that a relatively small driver amplitude produces a large growth in energy.

The driver effectiveness on the bottom boundary is plotted as a function of $x$ in Figure \ref{fig:poynting_flux}. The plot clearly shows the locations of the resonating field lines, where most of the Poynting flux is concentrated. Figure \ref{fig:poynting_flux} confirms that near the resonating field lines the driver effectiveness is greater than unity and away from the resonant lines it is much less than unity. The position of the resonant field line varies as the amplitude of the driver increases, due to nonlinear effects (see Section \ref{sec:nonlinearities}). The total energy in the domain is shown in Figure \ref{fig:total_energy} where the step-like profile corresponds to the period of the driver.

Figure \ref{fig:alfven_energy} shows that the total energy associated with the Alfv\'en waves, $E_A$, grows to a maximum, after which the energy oscillates about its time-averaged value (about 0.12$E_{lin}^{end}$). The Alfv\'en waves oscillate in the $z$-direction and so the energy density of an Alfv\'en wave, $e_A$, at a point in space is given by
$$
  e_A=\frac{1}{2}\rho v_z^2 + \frac{B_z^2}{2\mu}.
$$
The total Alfv\'en wave energy in the domain, $E_A$, is then given by
\begin{equation}
    E_A = \int_Se_AdS,
\end{equation}
where $S$ is the computational domain. The magneto-acoustic energy, $E_{acoustic}$, is defined to be the magnetic and kinetic energy associated with all perturbations which are not Alfv\'en waves and is defined as\begin{equation}
    E_{acoustic}=\int_S \frac{1}{2}\rho \left(v_x^2+v_y^2\right) + \frac{B_x^2+B_y^2}{2\mu}dS.
\end{equation}
Although the Alfv\'en wave energy stops growing (see Figure \ref{fig:alfven_energy}), the total energy of the system continues to grow (see Figure \ref{fig:total_energy}) until $t=20\tau_{driv}$ when the driver is switched off. Therefore, a steady state is reached, where all the energy generated by the time average of the Poynting flux goes into thermal energy and magneto-acoustic energy. In Section \ref{sec:discussion} it is shown that most of the energy goes into heat and not magneto-acoustic energy. Since a steady state is reached, Figure \ref{fig:poynting_flux} also gives a good indication of which field lines are heated most. From Figure \ref{fig:alfven_energy}, it can be seen that steady state is reached at about $t=5\tau_{driv}$ in the linear experiment (blue curve). It is interesting to note that the transition from a transient state to a steady state does not have a noticeable impact on the total energy evolution (see Figure \ref{fig:total_energy}) in the linear experiment.

From Figure \ref{fig:total_energy}, it can be seen that the growth of the total energy is linear. This is not surprising during steady state, as the amplitudes of the waves have stopped growing and so all the Poynting flux is transferred into heat at a constant rate. However, even during the transient phase ($t < 5\tau_{driv}$), the energy growth is linear. The linear growth during the transient phase occurs because the amplitude of the resonant wave grows quadratically with time, whereas the width of the resonant region decreases linearly with time (in accordance with resonant absorption theory \citep{Ionson1978}). Hence, the total energy growth is linear.

\subsection{Location of the heating}
\label{sec:location_of_heating}

\begin{figure}
    \centering
    \includegraphics[width=.49\textwidth]{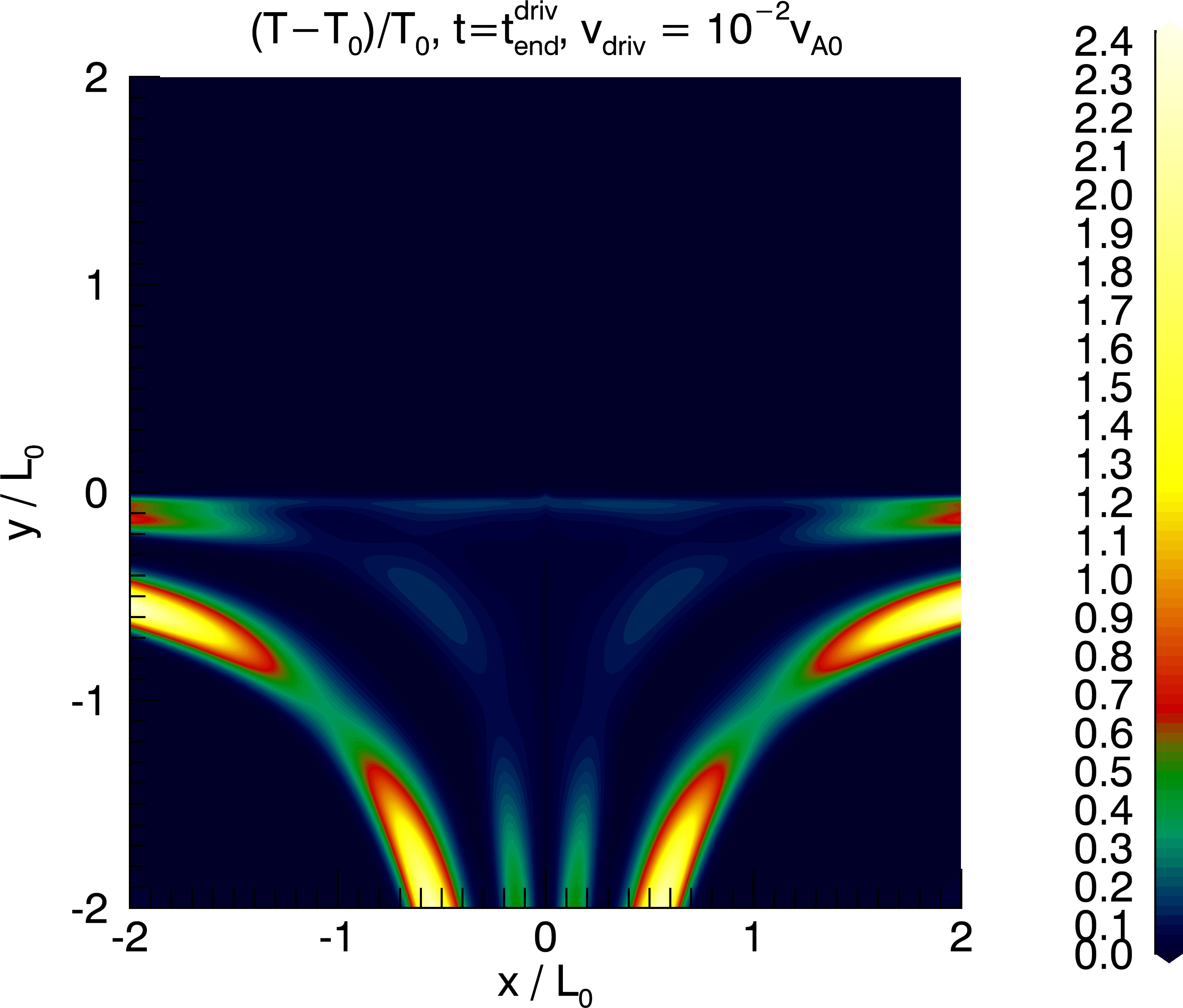}
    \caption{Contour plot showing the change in temperature relative to the initial temperature at $t=t_{end}^{driv}$.}
    \label{fig:heating_location}
\end{figure}

Most of the heating occurs at the nodes of the resonating standing field lines and this can be seen in Figure \ref{fig:heating_location}. As stated in Section \ref{sec:equations}, there is no viscous dissipation (besides shock viscosity) and therefore most of the heating occurs from Ohmic heating. Since most of the heating is generated by gradients in $B_z$ perpendicular to the magnetic field this means that most of the heating occurs where $B_z$ is largest. For standing Alfv\'en waves, the magnetic field component of the Alfv\'en wave is largest at the nodes of the standing wave and so most of the heating occurs at the nodes of the standing waves. Conversely, viscous dissipation acts on gradients in velocity and this would lead to heating occurring at the antinodes. However, \citet{VanDoorsselaere2007} showed from observational evidence that heating occurs mainly at loop footpoints and from this, they inferred that resistive heating dominates over viscous heating for wave heating mechanisms.

Perhaps unexpectedly, significantly more heating occurs near $(x,y)/L_0=(\pm2,0)$ compared with $(x,y)/L_0=(0,-2)$. A similar phenomenon has been shown in, for example, \citet{McLaughlin2013}. The author showed that the wavefronts of Alfv\'en waves which are not reflective remain planar as they approach the null and the current builds up exponentially with time, causing most of the heating to occur at the horizontal separatrices and furthest from the null.

\section{Nonlinear aspects}
\label{sec:nonlinearities}

The effects of nonlinear Alfv\'en waves have been researched extensively; see for example \citet{Verwichte1999} for details on Alfv\'en waves in 1D and \citet{Thurgood2013ponderomotive} for details in 2D and \citet{Terradas2004} for details on standing Alfv\'en waves. 
We analyse the effects of nonlinear density structures and investigate how the damping rate is affected by the nonlinearities.

\subsection{Density structures}
\label{sec:density_strucutres}

\begin{figure}
    \centering
    \includegraphics[width=0.49\textwidth]{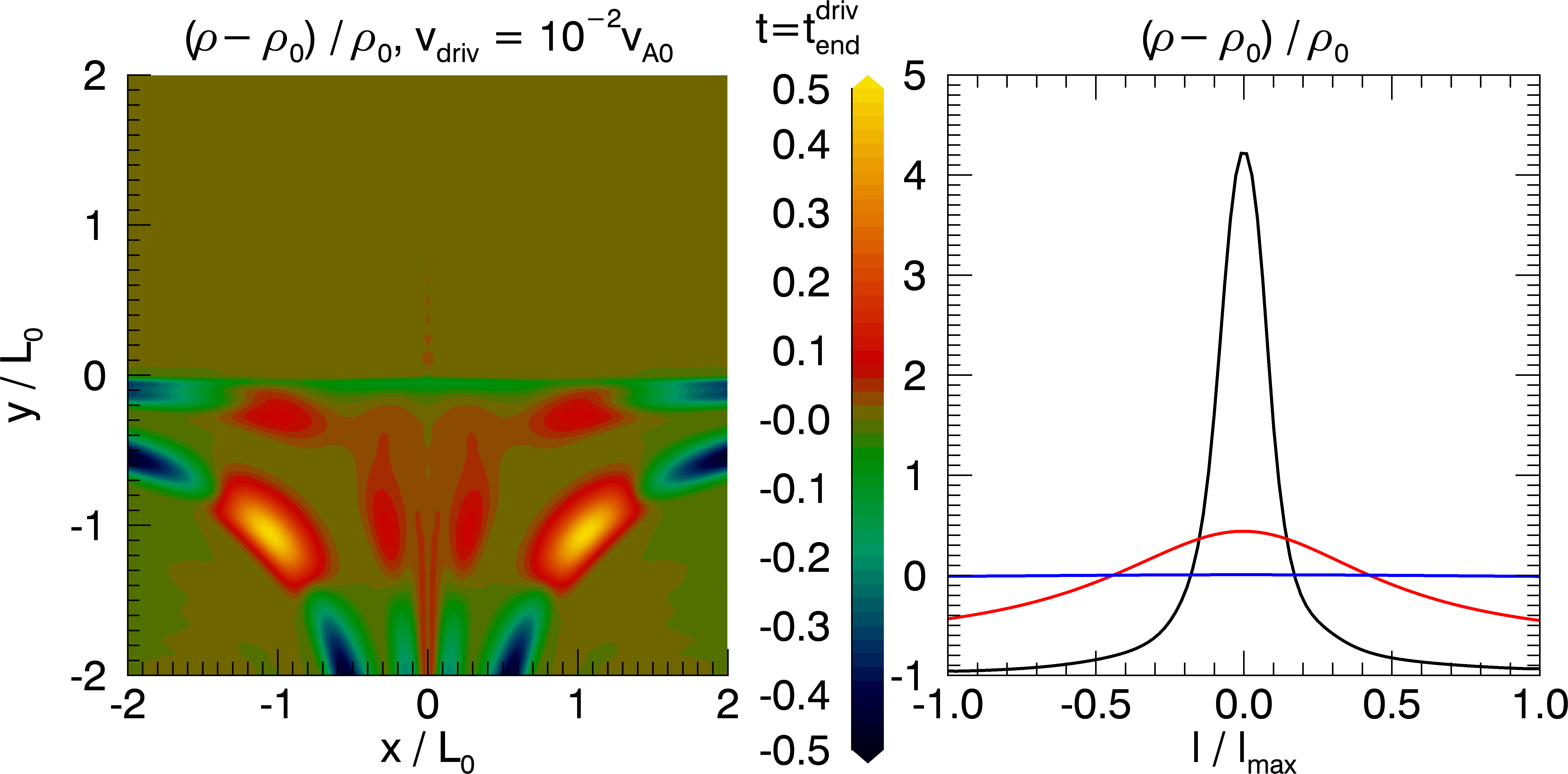}
    \caption{Left-hand plot: Contour of the density after the driving has finished for the experiment where a driver amplitude of $v_{driv}=10^{-2}v_{A0}$ is used. The right-hand plot shows the density along one of the outer most resonant field lines for all three experiments, where the same colour scheme is used as in previous plots.\ The value $l_{max}$ is equal to twice the length of the field line and $l$ is a variable giving the distance from the centre of the field line.}
    \label{fig:density_structures}
\end{figure}

\begin{figure}
    \centering
    \includegraphics[width=0.49\textwidth]{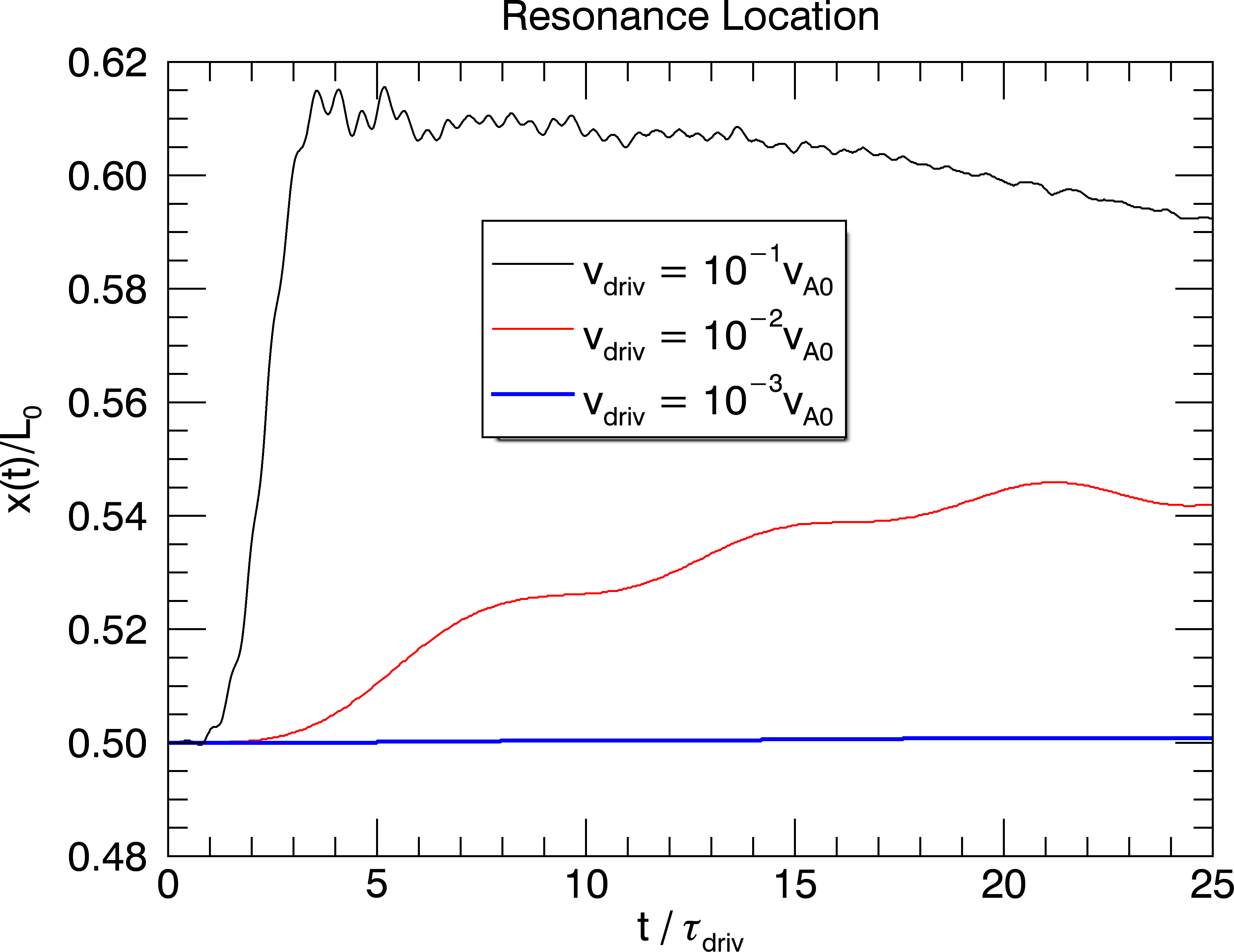}
    \caption{Plot of the $x$-coordinate of where the resonant field lines crosses the bottom boundary as a function of time (normalised by the period of the driver). The resonant location was calculated by finding the field line with a fundamental time period, given by Equation \eqref{eq:time_period_expression}, which is closest in value to the driving time period.}
    \label{fig:resonance_location}
\end{figure}

The nonlinearities generate density structures which can be seen in Figure \ref{fig:density_structures}, where red/yellow indicates density enhancement and green/blue indicates density reduction in the contour plots. Density structures are most pronounced along the resonating field lines where standing waves have been established. The density is largest at the antinodes of the standing waves and smallest at the nodes for two reasons. Firstly, as we only consider resistive heating and do not include viscosity, most of the heating takes place at the nodes. This causes the plasma pressure to increase at the nodes and hence a plasma pressure force is set up which pushes plasma away from the nodes towards the antinodes. If viscosity was included, it is likely some heating would also occur near the loop apex, reducing this effect. The second reason is that $B_z$ has its maximum amplitude at the nodes and smallest amplitude at the antinodes. This means that the nonlinear magnetic pressure force/ponderomotive force, $\vec{\nabla}B_z^2/(2\mu)$, also acts to push plasma towards the antinodes, away from the nodes. The right-hand side of Figure \ref{fig:density_structures} shows the density along one of the resonating field lines which are oscillating at the fundamental harmonic. It can be seen that plasma has been concentrated towards the apex/antinode of the field line and moved away from the footpoints/nodes of the field lines. For a more detailed analysis of the density structures formed by standing Alfv\'en waves, see \citet{Terradas2004}. These authors show that the amplitude of the density structures is proportional to $v_A/v_s$, where $v_s$ is the sound speed. Hence, if a higher plasma-beta were used then the amplitude of the density structures would be reduced.

One key effect these density structures have is that they cause the natural periods of the field lines to change, which in turn changes the location of the resonance. Indeed, in Figure \ref{fig:poynting_flux}, it can be seen that the peaks have shifted away from the origin for higher amplitude drivers. The density structures result in the natural periods of the field lines increasing. As the Alfv\'en speed is now a function of position along the field lines, even in a field-aligned coordinate system, the period is not simply given by the wavelength divided by the wave speed. The density structures have an enhancement in density at the antinodes and as the amplitude of the plasma velocity is highest at the antinodes, changes to the Alfv\'en speed at the antinodes affect the period the most. Since the density is enhanced at the antinodes, this results in a decrease in Alfv\'en speed at the antinodes and thus an increase in the period.

More rigorously, this result can be derived by considering the wave equation (see Appendix \ref{sec:harmonic_time_periods})
\begin{equation}
    \label{eq:wave_equation2}
    \frac{\partial^2v_z}{\partial t^2}=\frac{B_0^2}{\mu L_0^2}\frac{1}{\rho(s)}\frac{\partial^2 v_z}{\partial s^2},
\end{equation}
where $s$ is related to the distance along a field line. For simplicity, the density is assumed to be a function of only space and not time. In \citet{Terradas2004}, it was shown that for a standing Alfv\'en wave with angular frequency, $\omega$, given by
$$\omega=kv_A$$
then the density structures will oscillate with a frequency, $\omega_s$, which is approximately given by
$$\omega_s=2v_sk,$$
where $k$ is the wave number of the standing Alfv\'en wave. Hence, the simplification that the density is constant in time can be made if $v_s\ll v_A$ as this means $\omega_s\ll\omega$. A second simplification is made by assuming the time dependence of $v_z$ is given by $e^{i\omega t}$. By multiplying through by $v_z$, replacing time derivatives with $i\omega$ and using integration by parts, Equation \eqref{eq:wave_equation2} can be written as
\begin{equation}
    \omega^2=\frac{B_0^2}{\mu L_0^2}\frac{\int\left(\partial v_z/\partial s\right)^2ds}{\int\rho(s)v_z^2ds},
\end{equation}
provided the integrals are not taken at a time when the denominator is equal to zero. Hence,
\begin{equation}
    \label{eq:time_period_expression}
    \tau=2\pi\frac{L_0}{v_{A0}}\sqrt{\frac{\int\rho(s)v_z^2ds}{\int(\partial v_z/\partial s)^2ds}}.
\end{equation}
Equation \eqref{eq:time_period_expression} confirms that the value of the density near where the velocity is largest (the antinode) is the most important. Figure \ref{fig:resonance_location} shows an approximation of the location of the field line (in terms of the $x$-coordinate of where the field line crosses the bottom boundary) with a fundamental harmonic period which is closest in value to the period of the driver. In other words, the figure shows the approximate location of the fundamental harmonic resonant field line. We note that only the coordinates in the $x>0$ side of the domain are shown. For the nonlinear experiment (black line), Figure \ref{fig:resonance_location} predicts the location of the resonance to be at about 0.6$L_0$ compared to the actual location of about $0.7L_0$ (see Figure \ref{fig:poynting_flux}). Given the number of approximations, this is a reasonable agreement. In addition, Figure \ref{fig:resonance_location} gives an indication of how quickly and in which direction the resonance moves. To derive the resonance location more rigorously, the wave equation would need to be solved as a Sturm-Liouville problem and the eigenvalues which satisfy the boundary conditions would give the resonant frequencies. 

From Figure \ref{fig:total_energy}, it can be seen that the nonlinearities cause the system to be less effective at extracting energy from the driver. This is demonstrated by the fact that the final value for $(E_{tot}-E_{tot0})/E_{lin}^{end}$ is lower when a higher amplitude driver is used. One of the reasons the nonlinearities reduce the driver effectiveness is because they cause the resonance to shift location. Resonant field lines are effective at extracting energy from the driver because they generate a build-up in $B_z$, which increases the Poynting flux. Shifting the resonance location results in energy at the previous resonance location being lost owing to destructive interference. Moreover, the nonlinearities reduce the density at the footpoints, reducing the energy associated with the Alfv\'en waves generated by the boundary driver, resulting in less energy entering the system.

In addition to generating density structures, the ponderomotive force also acts to create a similarly shaped temperature profile. However, as shown in Section \ref{sec:location_of_heating}, the Ohmic heating is the dominant effect which determines the shape of the temperature profile in the experiments. In experiments where $\eta\rightarrow0$, there is no longer any Ohmic heating and the ponderomotive force now enhances the temperature at the antinodes owing to plasma compression.

\subsection{Damping rate}
\label{sec:damping_rate}

\begin{figure}
    \centering
    \includegraphics[width=0.49\textwidth]{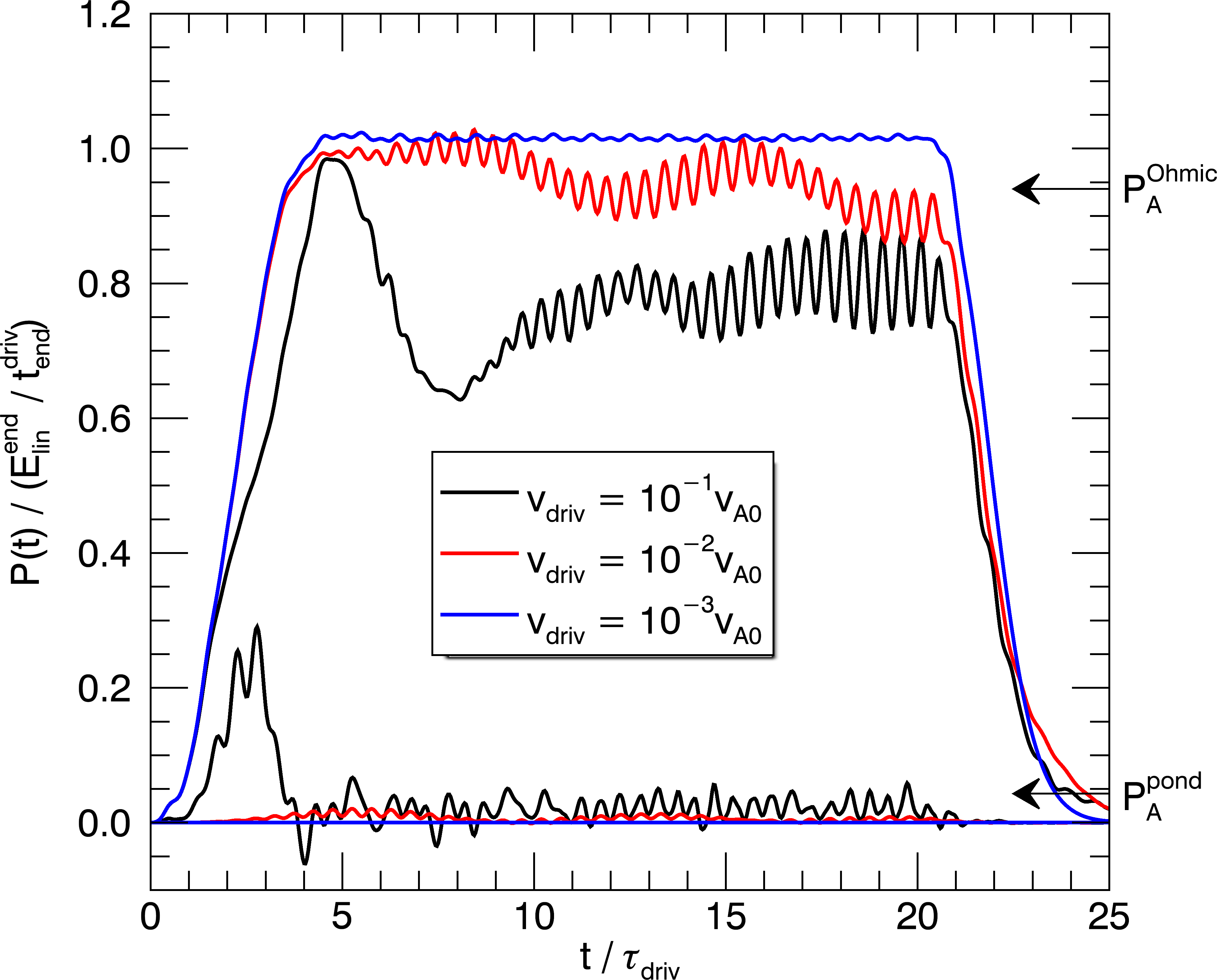}
    \caption{Plot of the Ohmic power, $P_A^{Ohmic}$, and the ponderomotive power, $P_A^{pond}$ as functions of time, for different driver amplitudes. Labels are provided on the right-hand side of the figure. The plots have been normalised by $E_{lin}^{end}/t_{end}^{driv}$, which gives the average power input from the driver in an equivalent but linear and ideal set-up.}
    \label{fig:ohmic_vs_pond_power}
\end{figure}

The term damping rate refers to the rate at which Alfv\'en wave energy is converted into other forms of energy.
The change in total Alfv\'en wave energy is given by
\begin{equation}
    \tag{\ref{eq:zenergy_evolution}}
    \frac{dE_A}{dt}=\frac{dE_{tot}}{dt}+\int_S\vec{v}\cdot\vec{\nabla}\left(\frac{B_z^2}{2\mu}\right)-\frac{1}{\sigma}\left(\frac{\vec{\nabla}B_z}{\mu}\right)^2dS,
\end{equation}
where Equation \eqref{eq:zenergy_evolution} is derived in Appendix \ref{sec:zenergy}. Equation \eqref{eq:zenergy_evolution} shows that in the absence of a driver there are two terms which affect the energy evolution of the Alfv\'en wave energy. The first term,
\begin{equation}
    \label{eq:ponderomotive_power}
    P_A^{pond}(t)=-\int_S\vec{v}\cdot\vec{\nabla}\left(\frac{B_z^2}{2\mu}\right)dS,
\end{equation}
is referred to as the ponderomotive power, which is related to the work done by the magnetic pressure force/ponderomotive force. The second term,
\begin{equation}
    P_A^{Ohmic}(t)=\frac{1}{\sigma}\int_S\left(\frac{\vec{\nabla}B_z}{\mu}\right)^2dS,
\end{equation}
is referred to as the Alfv\'en Ohmic power and is related to the Ohmic heating. The
ponderomotive power can both increase or decrease the Alfv\'en wave energy whereas the direct effect of the Ohmic heating only ever acts to reduce the Alfv\'en wave energy. It is worth noting that Equation \ref{eq:zenergy_evolution} shows that it is possible for flows perpendicular to the invariant direction to increase the amplitude of the Alfv\'en waves, however, it is impossible for the flows to change the amplitude if an Alfv\'en wave initially has zero amplitude.

Figure \ref{fig:ohmic_vs_pond_power} shows both the ponderomotive and Ohmic powers as functions of time. It can be seen that the net effect of the ponderomotive power is to increase the damping rate, however, its contribution is small compared to the Ohmic power. The ponderomotive power only has a large contribution at $t=2.5\tau_{driv}$. As stated in Section \ref{sec:density_strucutres}, \cite{Terradas2004} showed that for a 1D ideal standing  Alfv\'en wave which is not driven, the amplitude of the density structures oscillates between a maximum and minimum at an angular frequency given by
$$\omega_s=2v_s k,$$
where $k$ is the wave number of the Alfv\'en wave. These authors showed that the longitudinal velocity also oscillates with this frequency. Hence, by energy conservation, if the energy of the longitudinal flows oscillates, the energy of the Alfv\'en waves must also oscillate and this energy change is carried out by the ponderomotive power. Therefore, the initial increase and then decrease in the ponderomotive power up to $t = 3.5 \tau_{driv}$ (see the black curve in Figure \ref{fig:ohmic_vs_pond_power}) can be understood as the magnetic pressure force generating longitudinal flows and then the restoring plasma pressure force acting to return the density structures to their equilibrium position. After the initial peak in ponderomotive power, the power becomes negative around $t = 4 \tau_{driv}$, reflecting the fact that the plasma pressure force is acting to push the density structures back to their equilibrium position. However, by this point, the system has reached steady state and so now there is sufficient Ohmic heating for the associated pressure forces to dominate the motion of the longitudinal flows. As stated in Section \ref{sec:density_strucutres}, the pressure forces associated with the Ohmic heating act with the magnetic pressure force to push plasma away from the nodes. Hence, once steady state is reached, the ponderomotive power rarely becomes negative and oscillates with a period which is half that of the driving period.

Figure \ref{fig:ohmic_vs_pond_power} shows that the Ohmic power dominates over the ponderomotive power. One reason this property holds is that the ponderomotive power is a fourth-order nonlinear term because the velocity perturbations in the plane are second order, whereas the Ohmic power is a second-order nonlinear term. The Ohmic heating depends on the conductivity, $\sigma$, and so in a highly conductive plasma, it may at first seem that the ponderomotive power should dominate. However, if the plasma is highly conducting this often results in very short length scales forming and the Ohmic heating is inversely proportional to the square of the length scales while the ponderomotive power is only inversely to proportional to the length scale. Hence, we suggest that this property is also likely to hold in similar configurations.

Figure \ref{fig:alfven_energy} shows that in all the experiments, the Alfv\'en wave energy decays to zero at nearly the same rate when the driver is switched off. This is somewhat unexpected given that, in Figure \ref{fig:density_structures}, it can be seen that the nonlinearities generate large density structures and density structuring is usually associated with the enhancement of phase mixing and thus increased wave dissipation.  It would seem that because the density is only redistributed along the field lines in this figure, this does little to change the time taken for Alfv\'en waves to travel from one footpoint to another and so does little to enhance the phase mixing. 
In addition, the nonlinearities cause there to be an increase in coupling to compressive modes (\citet{Thurgood2013xpoint} and \citet{Thurgood2013ponderomotive}). As stated in the introduction, coupling to compressive modes has been proposed as an efficient mechanism for damping Alfv\'en waves. However, its contribution is relatively small; the coupling is a nonlinear effect and very quickly becomes negligible as the amplitude of the Alfv\'en wave decreases to zero. Evidence for this can be seen in Figure \ref{fig:ohmic_vs_pond_power}, where the ponderomotive power reaches zero much more quickly than the Ohmic power. For a more detailed analysis of the evolution of the compressive modes see \citet{Thurgood2013xpoint}, who study a nonlinear Alfv\'en pulse near an x-point with a similar magnetic field to that presented in this work.

\section{Discussion}
\label{sec:discussion}

\begin{figure}
    \centering
    \includegraphics[width=0.49\textwidth]{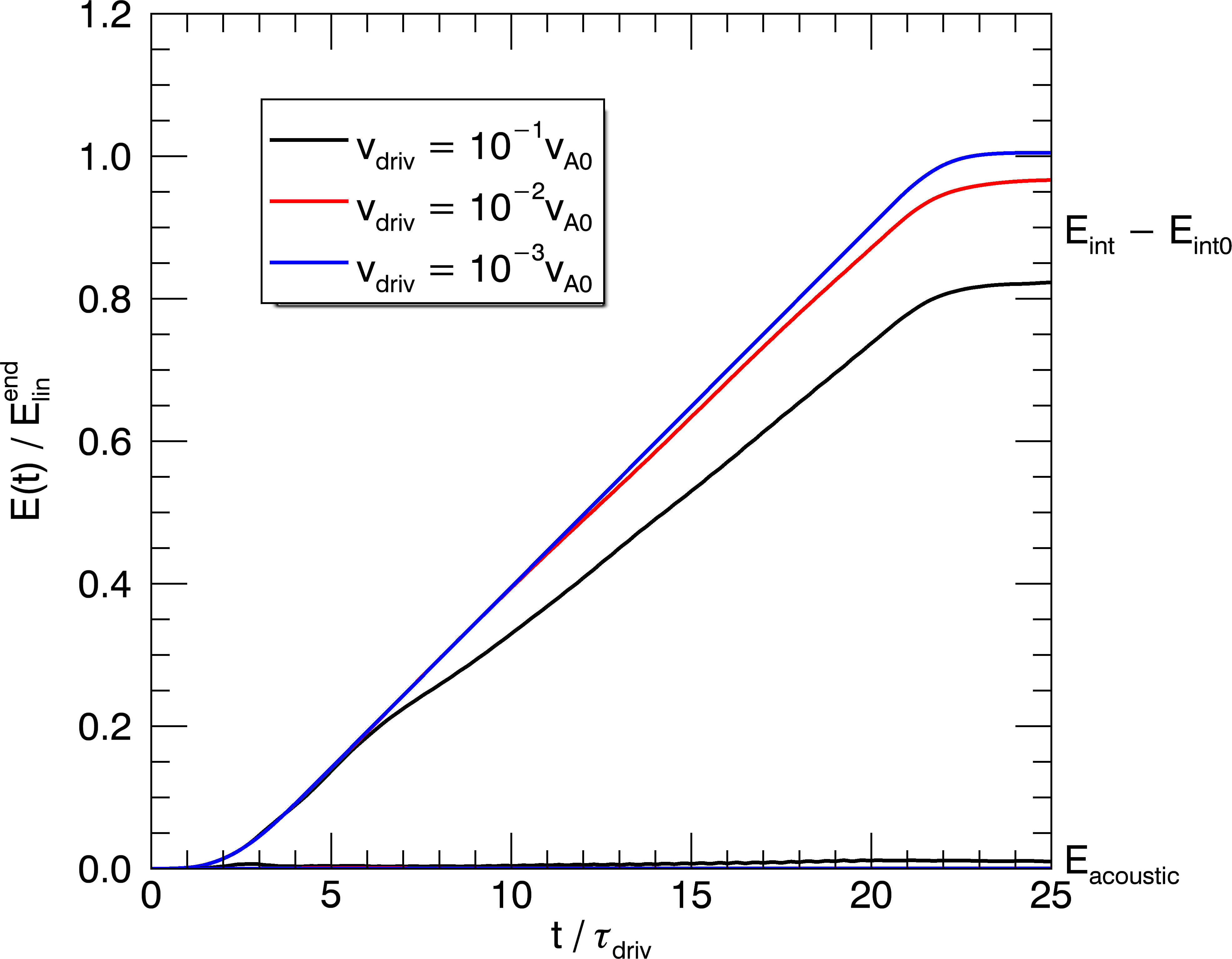}
    \caption{Plot of the internal and acoustic energy for different driver amplitudes. Labels are provided on the right-hand side of the figure. The plots have been normalised by $E_{lin}^{end}$, which gives the total energy input from the driver in an equivalent but linear and ideal set-up. }
    \label{fig:internal_and_acoustic_energy}
\end{figure}

\begin{figure}
    \centering
    \includegraphics[width=.49\textwidth]{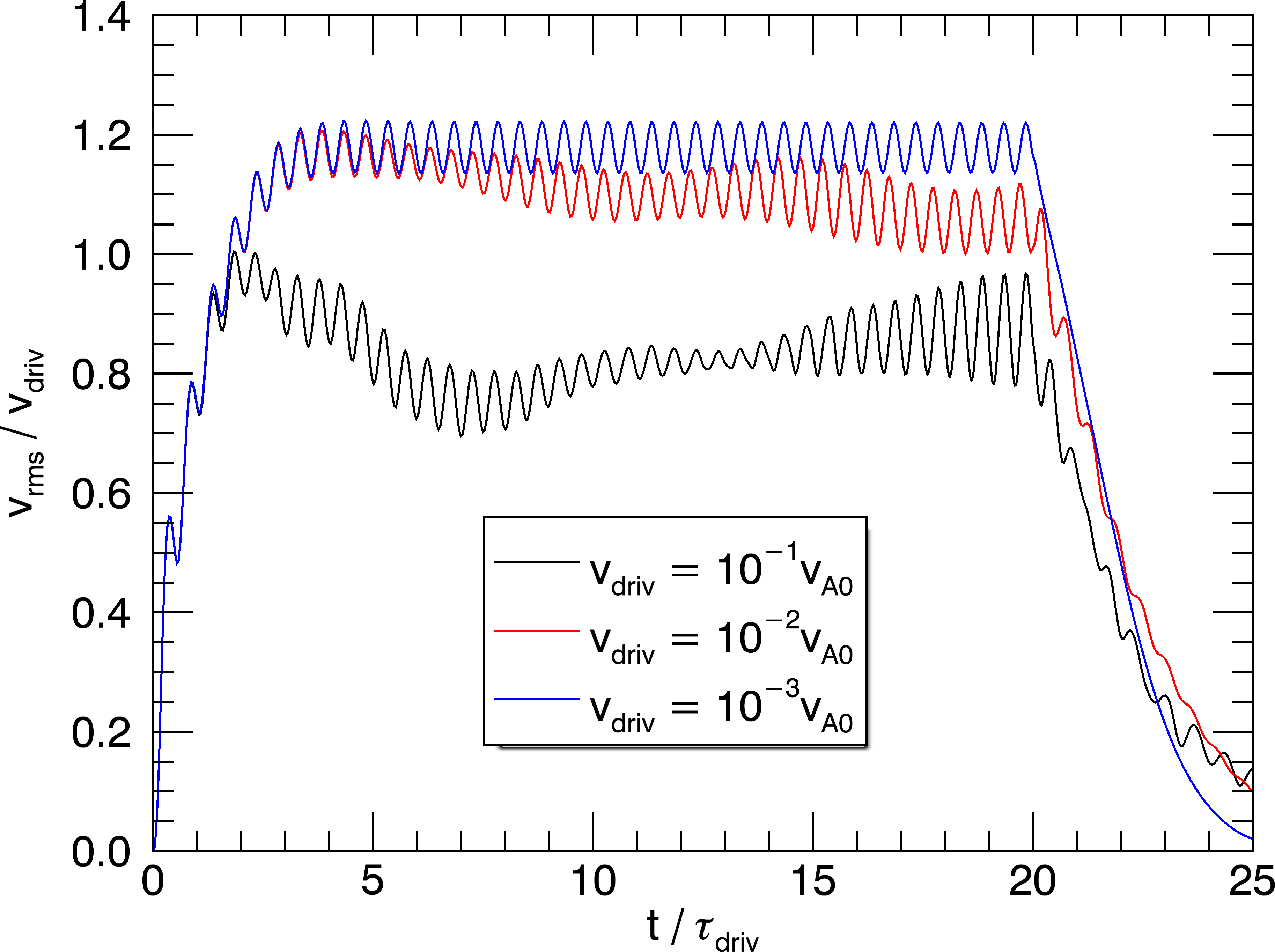}
    \caption{Plots of the ratio of the root mean square velocity in the bottom half of the domain to the driver amplitude.}
    \label{fig:average_v}
\end{figure}

\begin{table}
    \centering
    \begin{tabular}{| c | c | c | c |}
        \hline
         & Quiet & Coronal & Active \\
         & Sun & hole & region \\
        \hline
        Total coronal & & &  \\
        energy losses & $3 \times 10^2$ & $8 \times 10^2$ & $10^4$ \\
        (W\,$\text{m}^{-2}$) & & & \\
        \hline
    \end{tabular}
    \caption{Total coronal energy losses from conduction, radiation, and the solar wind in different regions of the corona, based on \citet{Withbroe1977}.}
    \label{tab:energy_losses}
\end{table}

This section aims to assess whether the model and driver presented in this work can provide sufficient heat to balance conductive and radiative losses in the quiet Sun. 
Let us assume that 100\% of the net energy provided by the driver (the Poynting flux) goes into heating the domain. The viability of this assumption is discussed at the end of this section. 

The average Poynting flux at a point along the driver boundary is given by
\begin{equation}
    \label{eq:average_poy_flux}
    \left\langle\frac{\vec{E}\times\vec{B}}{\mu}\right\rangle=v_{Ay}\rho_0v_{driv}^2\langle K_{driv}\rangle,
\end{equation}
for $t>\tau_{driv}/4$, where $\langle K_{driv}\rangle$ is defined as\begin{equation}
    \langle K_{driv}\rangle=\frac{1}{(x_{max}-x_{min})}\int_{x_{min}}^{x_{max}} K_{driv}(x)f(x)dx.
\end{equation}
In the experiments presented in this section, $\langle K_{driv}\rangle\approx$ 0.55, 0.53, 0.45 corresponding to $v_{driv}/v_{A0}=10^{-3},\ 10^{-2},\ 10^{-1}$, respectively. \citet{McIntosh2011} observed Alfv\'en waves with an average amplitude between $20-25$\,km\,$\text{s}^{-1}$ at a height of 15\,Mm in the quiet region of the solar corona. This velocity seems plausible, as the amplitude of an Alfv\'en wave scales with $\rho^{-1/4}$, provided there is no reflection. Photospheric motions are approximately $1-2$ km/s (\citet{Belien1999} and \citet{Moriyasu2004}), giving a velocity amplitude around 100 times larger at the top of the chromosphere. There is some reflection at the transition region, and therefore the value at the top of the transition region observed by \citet{McIntosh2011} is not unreasonable. From Figure \ref{fig:average_v} it can be seen that at steady state, the ratio of the average amplitude to driver amplitude is approximately unity. Therefore, if the following values (taken from \citet{McIntosh2011}) are used: $v_{driv}=20-25$\,km\,$\text{s}^{-1}$, $v_{Ay}=200-250$\,km\,$\text{s}^{-1}$, $\rho_0=(5-10)\times10^{-13}$\,kg\,$\text{m}^{-3}$, $\langle K_{driv}\rangle=0.5$, then the Poynting flux is given by
\begin{equation}
    v_{Ay}\rho_0v_{driv}^2\langle K_{driv}\rangle \approx 20 - 80\, \text{W}\, \text{m}^{-2}.
\end{equation}
At steady state, the wave energy stops growing and so 100\% Poynting flux provided by the driver goes into heat. The Poynting flux obtained above is of the order of the required flux to balance energy losses in the quiet Sun (see Table \ref{tab:energy_losses}), suggesting that phase mixing of Alfv\'en waves, as in this model, may indeed play a significant role in the heating of the corona. However, the model presented made many simplifications and the remainder of this section discusses the potential consequences of some of these simplifications. 

\begin{figure}
    \centering
    \includegraphics[width=0.49\textwidth]{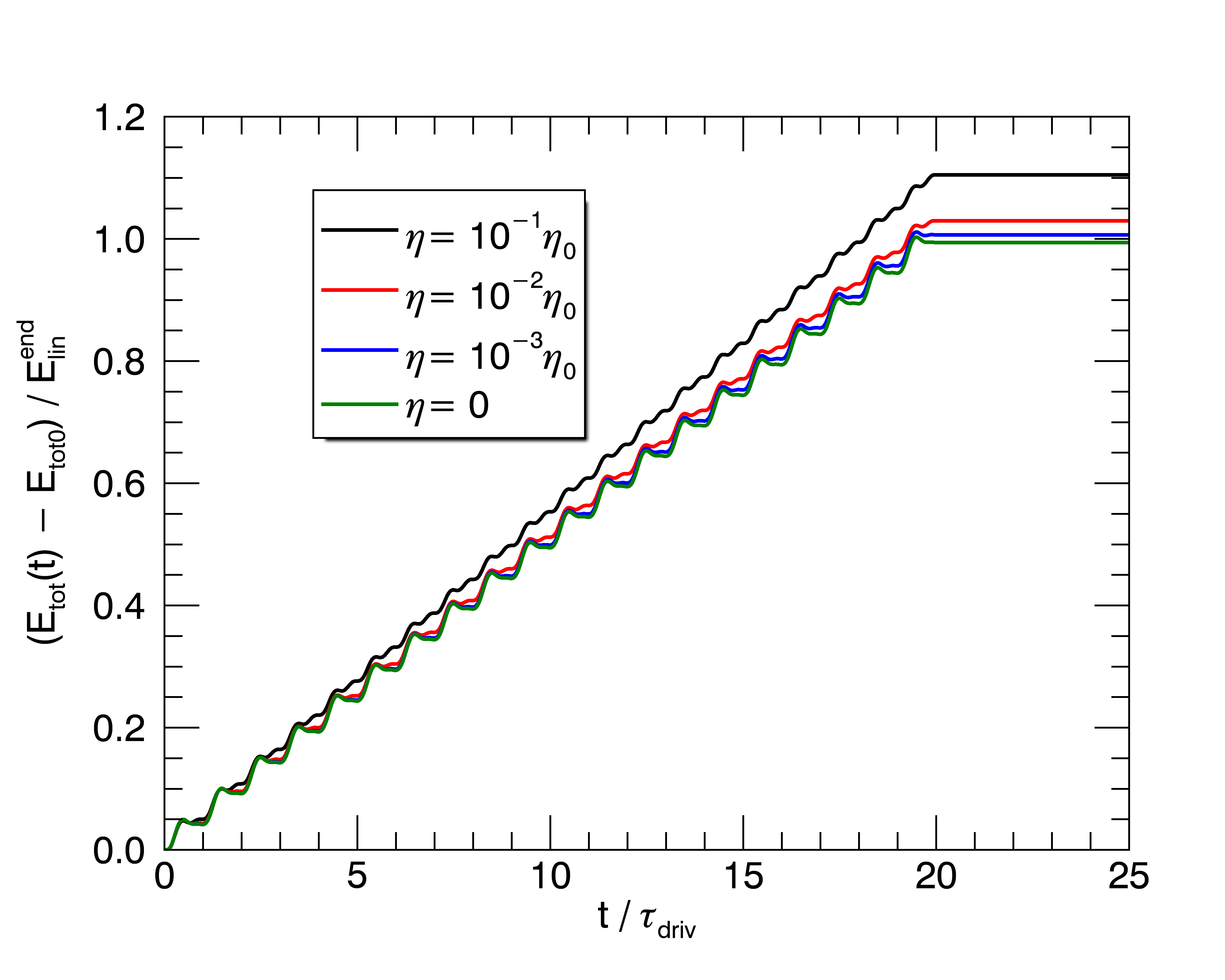}
    \caption{Plot of the total energy of the domain for different values of $\eta$. In the $\eta=0$ experiment there is no energy transfer due to Ohmic heating, however, there are still energy losses through numerical dissipation. The total energy was calculated using the Poynting flux on the boundary, therefore, any numerical energy losses in the domain are accounted for}. The plots have been normalised by $E_{lin}^{end}$, which gives the total energy input from the driver in an equivalent but linear and ideal set-up.
    \label{fig:high_vs_low_eta_total_energy}
\end{figure}

\begin{figure}
    \centering
    \includegraphics[width=0.49\textwidth]{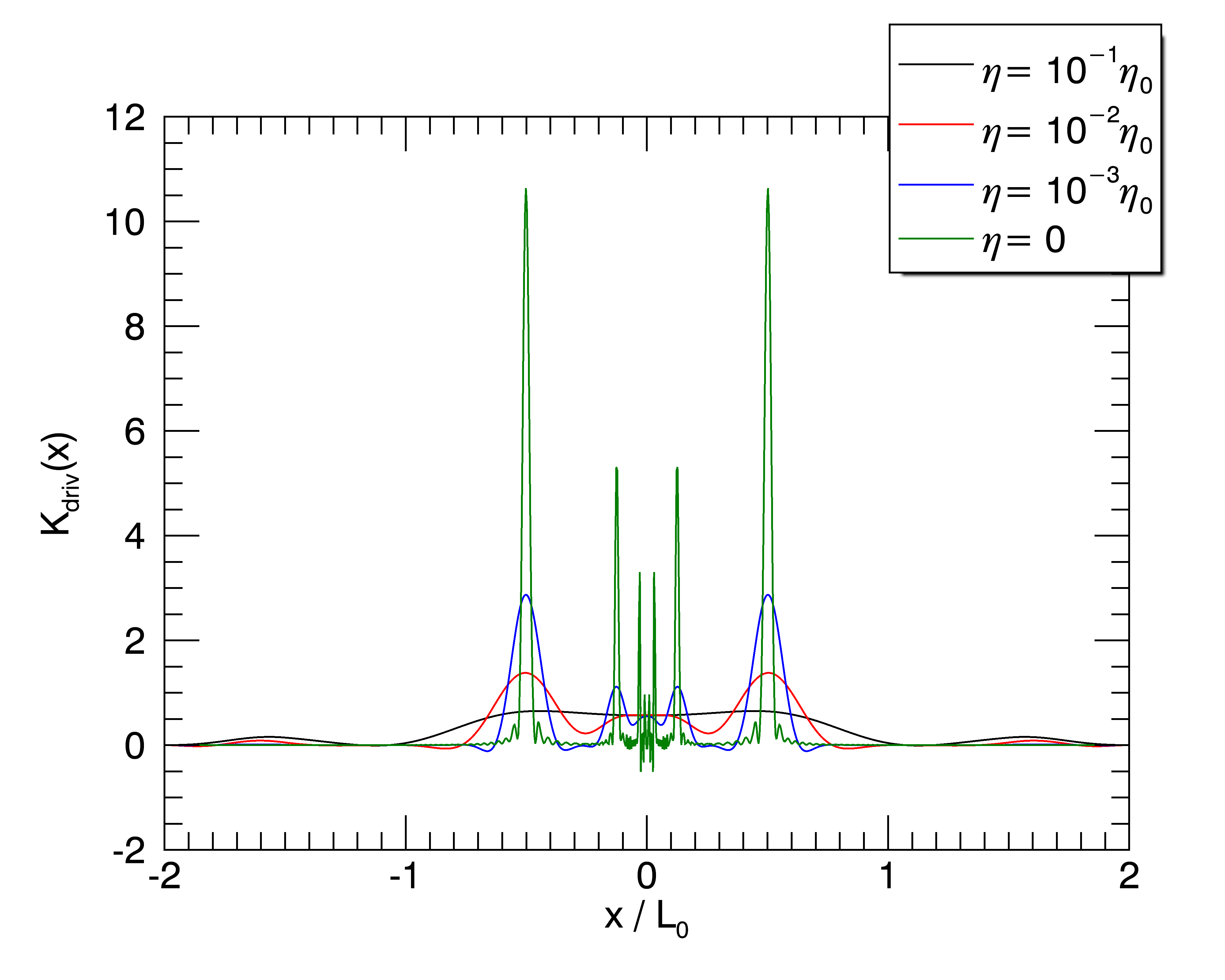}
    \caption{Plot of the driver effectiveness (Eq.~\eqref{eq:driver_effectiveness}) for different values of $\eta$. We note that numerical dissipation occurs in all the experiments including the $\eta=0$ experiment.} The plots have been normalised by $E_{lin}^{end}$, which gives the total energy input from the driver in an equivalent but linear and ideal set-up.
    \label{fig:high_vs_low_eta_poynting_flux}
\end{figure}


\begin{figure}
    \centering
    \includegraphics[width=0.49\textwidth]{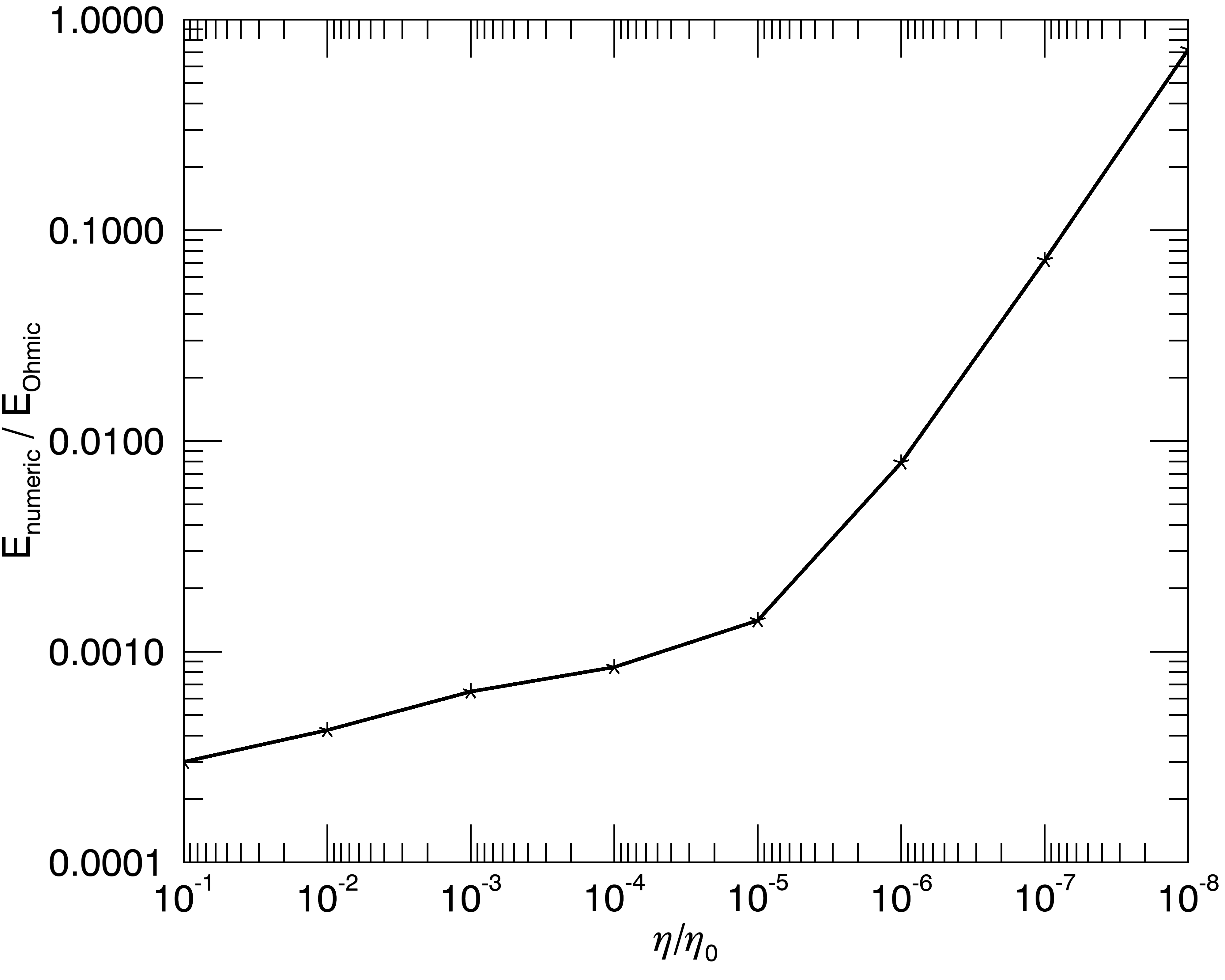}
    \caption{Plot of the ratio, $E_{numeric}/E_{Ohmic}$, for a range of $\eta$ values. The quantity $E_{numeric}$ refers to the energy in the domain which is lost through numerical diffusion and is estimated by comparing the total (volume integrated) energy in the domain with the Poynting flux through the driven boundary. The quantity $E_{Ohmic}$ gives the total amount of energy produced by Ohmic heating.}
    \label{fig:E_num_vs_E_ohmic}
\end{figure}

\begin{figure}
    \centering
    \includegraphics[width=0.49\textwidth]{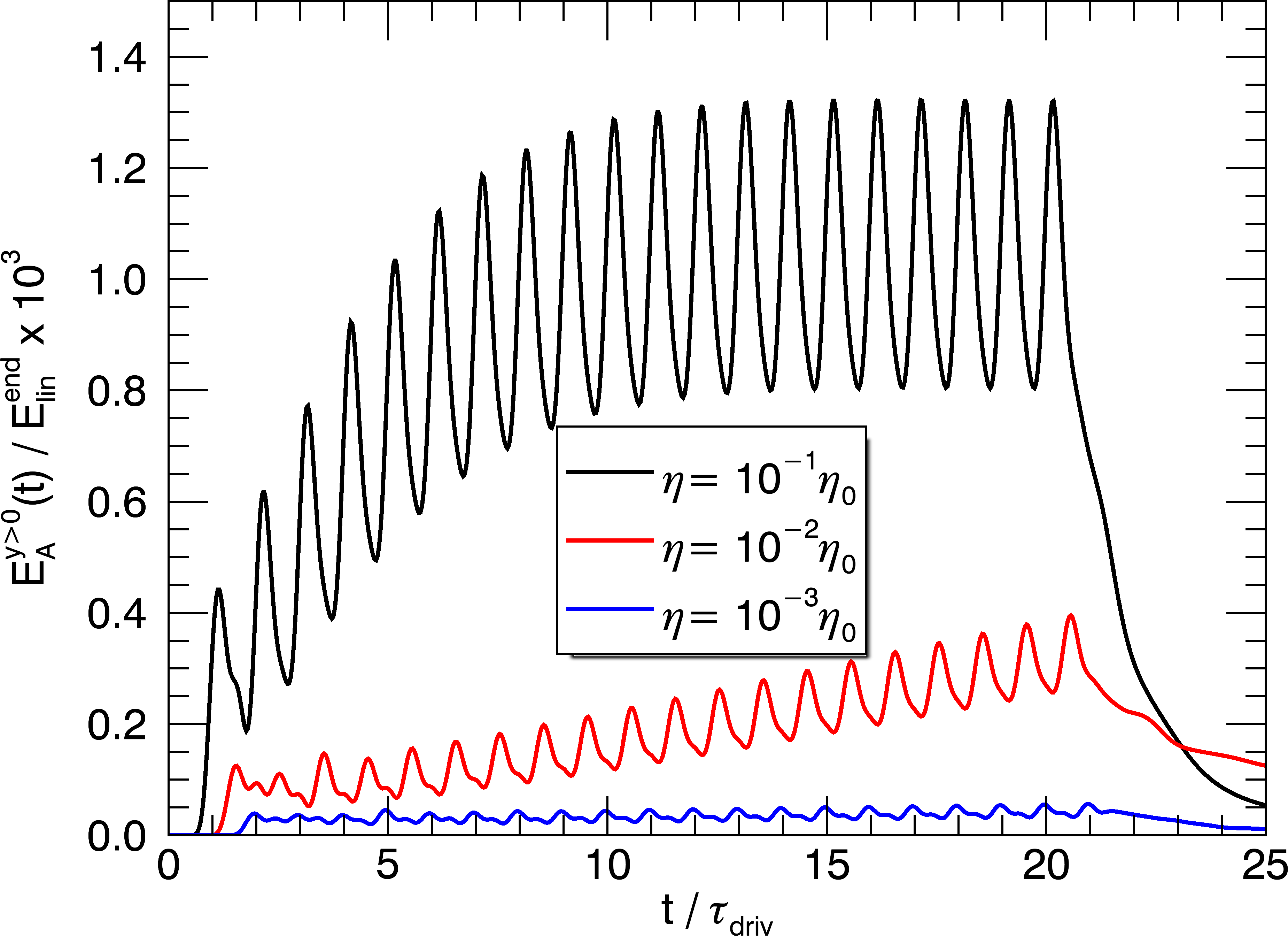}
    \caption{Plot of the Alfv\'en wave energy in the top half of the domain, $E_A^{y>0}$, for different values of $\eta$. The plots have been normalised by $E_{lin}^{end}$, which gives the total energy input from the driver in an equivalent but linear and ideal set-up.}
    \label{fig:alfven_energy_top_half}
\end{figure}

\begin{figure}
    \centering
    \includegraphics[width=.49\textwidth]{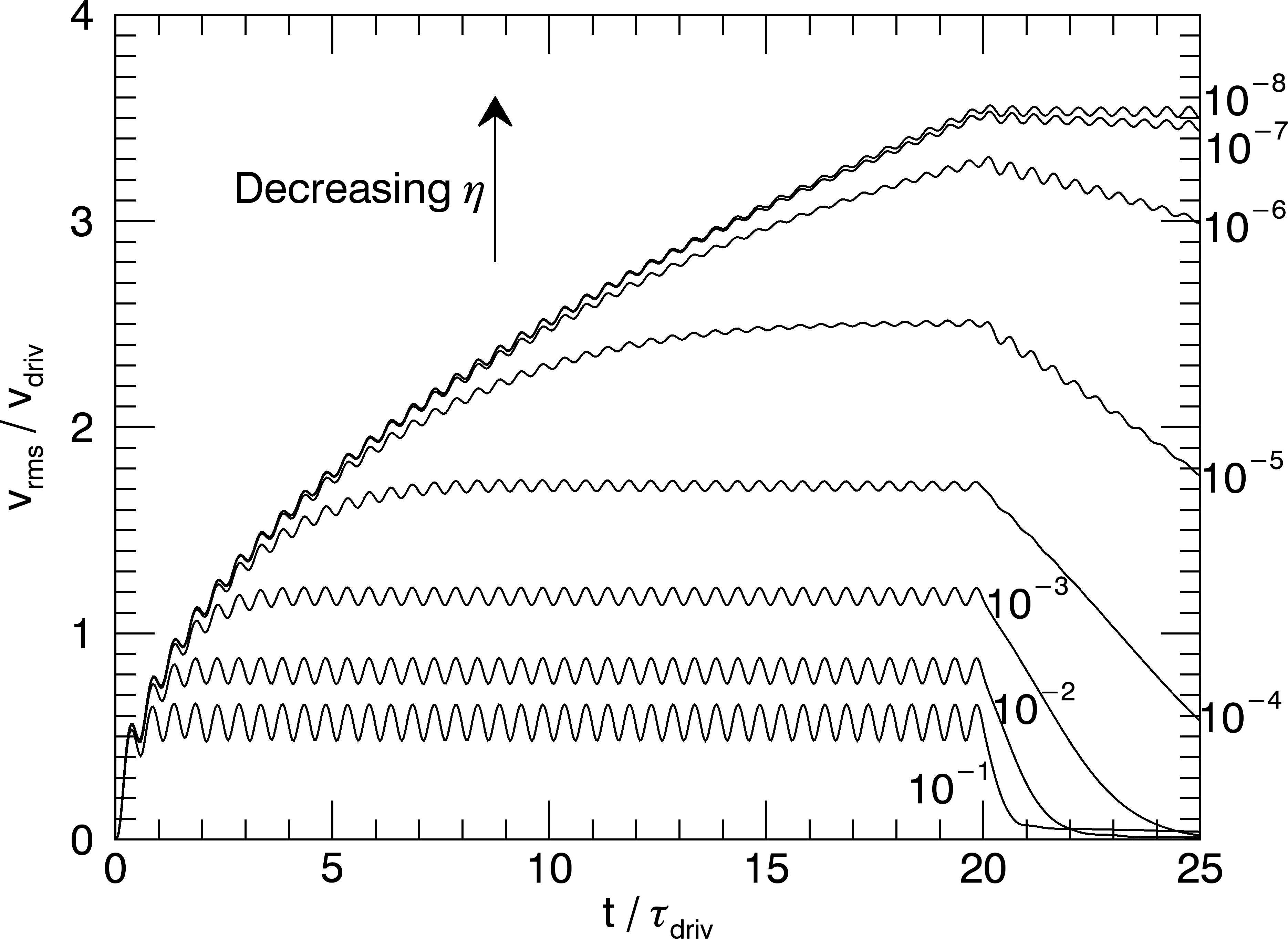}
    \caption{Plots of the ratio of the root mean square velocity to driver amplitude for different values of magnetic diffusivity. In this case, $\eta/\eta_0=[10^{-1},10^{-2},...,10^{-8}]$ with the bottom curve corresponding to $10^{-1}$ with each successive curve corresponding to the next value of $\eta$ in the list above. }
    \label{fig:high_vs_low_eta_average_v}
\end{figure}

\begin{figure}
    \centering
    \includegraphics[width=.49\textwidth]{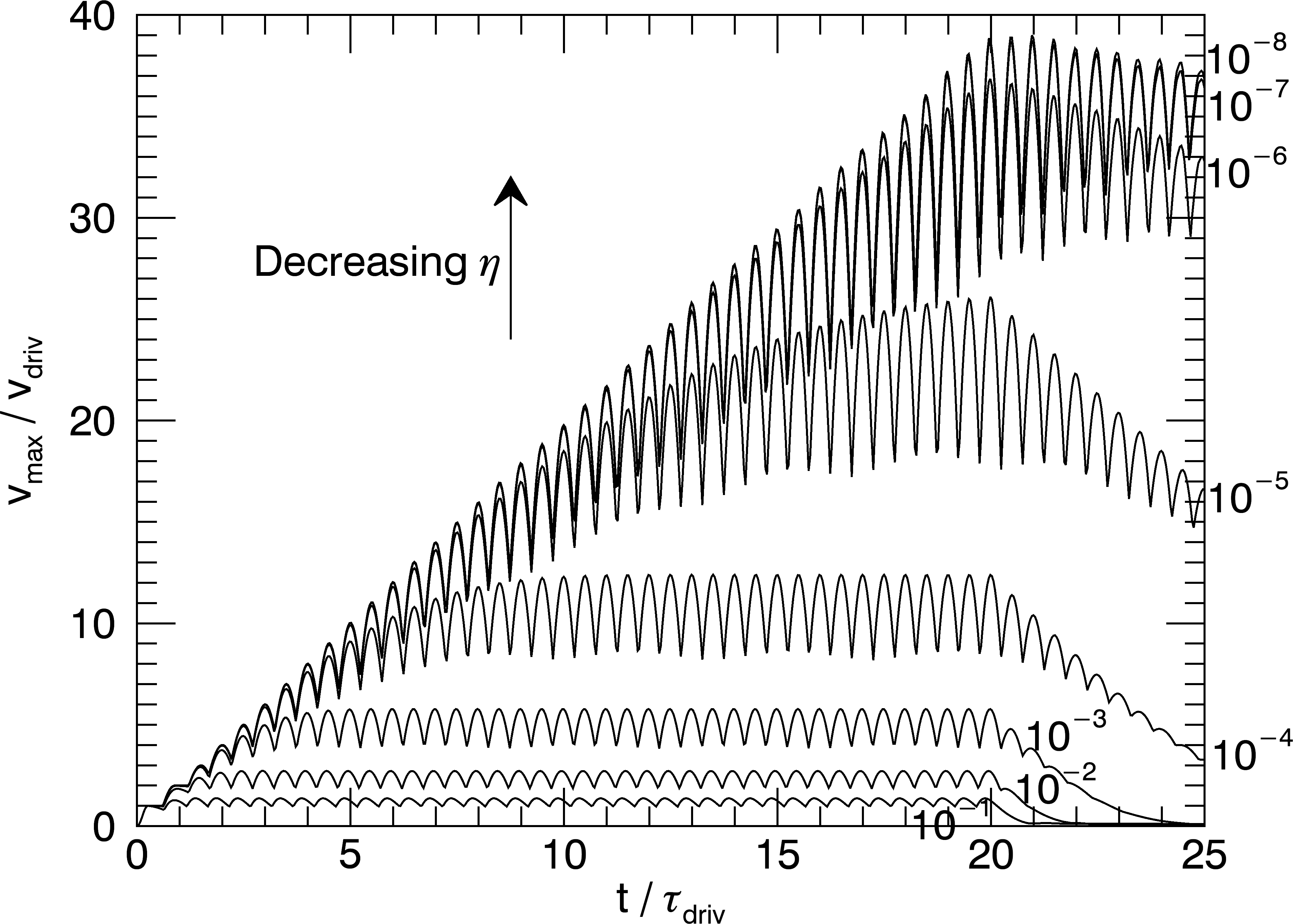}
    \caption{Plots of the maximum velocity in the domain for different values of magnetic diffusivity. In this case, $\eta/\eta_0=[10^{-1},10^{-2},...,10^{-8}]$ with the bottom curve corresponding to $10^{-1}$ with each successive curve corresponding to the next value of $\eta$ in the list above.}
    \label{fig:high_vs_low_eta_max_v}
\end{figure}

In Section \ref{sec:method}, it was shown that $\eta$ is too large by about a factor of $10^9$. Figure \ref{fig:high_vs_low_eta_total_energy} shows results from experiments where $\eta$ was varied. For these figures, the driver amplitude is set equal to
$$v_{driv}=10^{-3}v_{A}^{norm}\,,$$
and hence, the results are mostly linear. Also, the shock viscosity was switched off in these experiments such that the experiment can be considered close to ideal, although some amount of numerical diffusion is still present. We see that the total energy increases with $\eta$, but appears to converge towards a minimum as $\eta$ decreases. In the $\eta=0$ experiment, the total energy was calculated by calculating the amount of Poynting flux entering the system through the driver. For small $\eta$ (and for fixed $v_{driv}$), it appears the total energy evolution is independent of $\eta$, A similar phenomenon has been reported in the literature. For example, \citet{Wright1996} measured the total Ohmic heating in a similar phase mixing experiment and proved analytically that once steady state is reached, the total spatially integrated Ohmic dissipation is independent of $\eta$. There is a subtle difference between the result derived in \citet{Wright1996} and the result obtained in this work. The total energy evolution becomes independent of $\eta$ for small $\eta$ whereas \citet{Wright1996} showed that the total Ohmic heating is independent of $\eta$ once steady state is reached.  The proof in \citet{Wright1996} assumes that the simulation has run sufficiently long for steady state to be reached, however, in this paper, for the $\eta=0$ experiment, steady state is not reached during the simulation time. Figure \ref{fig:high_vs_low_eta_poynting_flux} helps to illustrate why the total energy converges to a limit for smaller $\eta$. The figure shows that increasing $\eta$ acts to increase the width of the resonating region, however, it also reduces the height and so there is little change to the total energy.

Even though there is no (explicit) magnetic diffusion in the $\eta=0$ experiment, wave energy is still dissipated through numerical diffusion. Figure \ref{fig:E_num_vs_E_ohmic} is presented such that the importance of numerical diffusion can be assessed. The figure shows the ratio of $E_{numeric}$ to $E_{Ohmic}$, where $E_{numeric}$ gives the energy in the domain which is lost through numerical diffusion and $E_{Ohmic}$ gives the total amount of energy produced by Ohmic heating. It can be seen that for $\eta>10^{-8}\eta_0$ Ohmic heating dominates and for this reason, we only considered experiments with $\eta>10^{-8}\eta_0$.


Increasing $\eta$ results in more Alfv\'en waves leaking across the separatrices (see Figure \ref{fig:alfven_energy_top_half}). None of the field lines in the bottom half of the domain enter the top half of the domain, hence, only a small fraction of the Alfv\'en waves leak into the top half of the domain. The Alfv\'en waves that travel into the top half of the domain do so via magnetic diffusion, which enables Alfv\'en waves to leak onto neighbouring field lines. Figure \ref{fig:alfven_energy_top_half} shows the amount of Alfv\'en wave energy in the top half of the domain, denoted by $E_A^{y>0}$, as a function of time. Figure \ref{fig:alfven_energy_top_half} shows that increasing $\eta$ results in more leakage, however when compared with Figure \ref{fig:high_vs_low_eta_total_energy}, it can be seen that even in the higher $\eta$ experiments the amount of leakage is still negligible when compared with the total energy increase of the system. Although nonlinearities appear to increase the amount of wave leakage, even in our most nonlinear experiments, the Alfv\'en wave energy which leaks across the separatrices is no more than 0.1\% (for $\eta=10^{-3}\eta_0$) and this tends to zero as $\eta\rightarrow0$.

Our earlier assessment of the Poynting flux was based on comparing $v_{rms}$ in our experiments with the values estimated by \cite{McIntosh2011} from SDO/AIA observations and assuming that $v_{driv} \sim v_{rms}$ (see Fig.~\ref{fig:average_v} for $\eta = 10^{-3}\eta_0$). However, for smaller values of $\eta$, the ratio $v_{rms}/v_{driv}$ increases as can be seen in Fig.~\ref{fig:high_vs_low_eta_average_v}. Therefore, to compare with the same, observed value of $v_{rms}$, we would have to reduce $v_{driv}$, resulting in a smaller Poynting flux. At the same time, for smaller values of $\eta$, nonlinear effects become more important as the maximum velocities in the domain grow substantially (see Fig.~\ref{fig:high_vs_low_eta_max_v}). \cite{McIntosh2011} argued that the observed amplitudes in the quiet Sun are of the order of 10\% of the local Alfv\'en speed. From Fig.~\ref{fig:average_v}, we can see that for amplitudes of this order ($v_{driv} = 10^{-1} v_{A0}$), $v_{rms}/v_{driv}$ is smaller than in the corresponding linear experiments. This can be explained by the fact that according to \citet{Verwichte1999} nonlinear damping mechanisms grow with $\sim(v/v_A)^2t$, where $v$ is the amplitude of a wave. Hence, we expect the effect of reducing $\eta$ on $v_{rms}/v_{driv}$ to be less significant than for the linear experiments shown in Fig.~\ref{fig:high_vs_low_eta_average_v}. Therefore, if observed values of $v_{rms}$ can be considered to be nonlinear, our assumption that $v_{driv} \sim v_{rms}$ is perhaps not unreasonable, even for smaller values of $\eta$.

Finally, we point out that the assumption that 100\% of the net energy provided by the driver goes into heating the domain depends on the frequency of the driver. For a constant frequency driver, the system eventually reaches a steady state where 100\% of the Poynting flux from the driver goes into heat. However, for a non-constant driver frequency, the system may be unable to reach steady state before the frequency profile of the driver changes, particularly, if a smaller value of $\eta$ were used. Using a random driver could reduce the driver effectiveness as changes to the driver frequency could lead to destructive interference with pre-existing waves. The effects of using a random driver imposed on a zero-dimensional harmonic oscillator are reviewed in \citet{Masoliver1993} and \citet{Gitterman2013}. A random driver in a 2D phase mixing experiment is investigated in \citet{Wright1995}. The effects of a random driver imposed on a 3D coronal loop are studied in \citet{DeGroof2002a} and \citet{DeGroof2002b}.

\section{Conclusions}
\label{sec:conclusions}

In this paper, we demonstrate that phase mixing can occur because of variations in field strength and field line length without the need for variations in density. The model deliberately does not impose any initial density structures as demonstrated by \citet{Cargill2016} that heating from phase mixing cannot sustain the density structures self-consistently.

We find that the nonlinearities reduce the driver effectiveness, which results in the total amount of heating being reduced.
For the range of driver amplitudes studied in this paper, the reduction in effectiveness is found to be about 20\%. Density structures are generated by the ponderomotive force and by pressure forces associated with the heating; this causes the resonance location to shift, which means energy build up is smaller than it would be otherwise. In addition to this, since the density at the boundary is reduced, the energy associated with the Alfv\'en waves entering the system is also reduced and so less energy enters the domain. The nonlinearities have a comparatively small effect on the damping rate (for the range of amplitudes studied in this work), where the damping rate is related to the rate at which the energy associated with the Alfv\'en waves is converted into other forms of energy.


We calculated an order of magnitude estimate of the Poynting flux to determine if the model presented in this work provides enough energy to balance conductive and radiative losses in the coronal region. We find that the Poytning flux provided in the model, with a large magnetic diffusion ($\eta=10^-3\eta_0$), indeed provides energy of the order necessary to balance conductive and radiative losses in the quiet Sun corona (but not active regions). We did not consider coronal holes because they are typically composed of open magnetic field lines and our model addresses closed loops. The order of magnitude estimate was constrained by ensuring that the steady-state, root-mean-square velocity, $v_{rms}$, matches observations \citep{McIntosh2011}. We show that as $\eta$ decreases, the Poynting flux remains approximately constant. However, $v_{rms}$ increases and this means that for smaller $\eta$ the driver amplitude must be reduced to ensure $v_{rms}$ remains fixed. We estimated from linear experiments that if a physical value of $\eta$ were used, the driver amplitude would have to be reduced by approximately a factor of 10, resulting in a decrease in Poynting flux by a factor of 100.
From Figure \ref{fig:average_v} it can be seen that nonlinearities reduce the root-mean-square velocity. One possible mechanism for this could be the nonlinear self-modification of Alfv\'en waves which, as shown by \citet{Verwichte1999}, results in the formation of strong currents and hence strong Ohmic dissipation in a time that is proportional to $(v/v_A)^{-2}$. Thus, although our linear results suggest that with a realistic value of $\eta$ the model does not produce enough heat to balance losses in the corona, it is still plausible that there might be enough heat in a nonlinear model.

It has long been known that Alfv\'en waves can mode convert to magnetoacoustic waves (as described in  \citet{Verwichte1999} and \citet{Thurgood2013ponderomotive}) and that the Alfv\'en waves generate density structures as shown in \citet{Terradas2004}. Equation \eqref{eq:zenergy_evolution} shows that the Alfv\'en waves can transfer energy into flows perpendicular to the invariant direction and vice versa by doing work through the magnetic pressure force (ponderomotive force). Therefore, if a large velocity is imposed in the longitudinal direction, this could result in large changes to the energy of the Alfv\'en wave.

\vspace{2.5mm}

\textbf{Acknowledgements.} This research has received funding from the Science and Technology Facilities Council (UK) through the consolidated grant
ST/N000609/1 and the European Research Council (ERC) under the European Union's Horizon 2020 research and innovation programme (grant agreement
No. 647214).

\begin{appendix}

\section{Ideal and linear analytic solution}
\label{sec:ideal_and_linear_solution_apdx}

The ideal and linear solution was calculated by solving the wave equation (Equation \ref{eq:wave_equation3}) using d'Alembert's formula for both $v_z$ and $B_z$. We note that the solution is a superposition of Heaviside functions which enter the domain owing to wave reflection. A simplification is made by assuming the driver is given by
$$v_z=v_{driv}\sin(\omega t),$$
with no ramp-up period involving the square of a sine (as used in Equation \eqref{eq:driver_time_profile}).
Once $B_z$ and $v_z$ are calculated, the Poytning flux on the boundary is calculated as
$$\begin{aligned}
-\frac{B_y}{\mu}v_zB_z=&v_{A}^y\rho_0v_{driv}^2\sin(\omega t)\left(\sin\left[\omega\left(t-\frac{2ml}{v_{A0}}\right)\right] \right. \\
+&\left.2\cot\left(\frac{\omega l}{v_{A0}}\right)\sin\left(\frac{m\omega l}{v_{A0}}\right)\sin\left[\omega\left(t-\frac{ml}{v_{A0}}\right)\right]\right),\\
\end{aligned}$$
for  $\omega l/v_{A0}\neq k\pi$, where $k$ is an integer, $l=l(x)$ is the length of the loop given by $s_2-s_1$ in Appendix \ref{sec:harmonic_time_periods}, $v_{Ay}=B_y/\sqrt{\rho_0\mu}$ and $m$ is an integer given by
$$m=\left\lfloor\frac{tv_{A0}}{2l}\right\rfloor\; .$$
The floor brackets, $\lfloor\; \rfloor $, correspond to the largest integer smaller than $tv_{A0}/(2l)$. The apparent singularity at $\omega l/v_{A0}=k\pi$, can be resolved and if $\omega l/v_{A0}=k\pi$ then the Poynting flux is given by
$$-\frac{B_y}{\mu}v_zB_z=v_{A}^y\rho_0v_{driv}^2(2m+1)\sin^2(\omega t).$$
This equation shows that the Poynting flux grows linearly with time along resonant field lines and so the energy grows quadratically. To calculate $E_{lin}^{end}$, the Poytning flux was then integrated along the bottom boundary in space and time, where $x$ goes from $x_{min}$ to $x_{max}$ and $t$ goes from $0$ to $t_{end}^{driv}$.

\section{Total energy evolution}
\label{sec:total_energy_evolution}
Using Equations \eqref{eq:continuity} - \eqref{eq:iduction} as well as Faraday's law, it can be shown that the rate of change of energy at a point in space in the domain is given by
\begin{equation}
    \label{eq:total_energy_equation}
    \frac{\partial}{\partial t}\left(\frac{p}{\gamma - 1}+\frac{B^2}{2\mu}+\frac{1}{2}\rho v^2\right) + \vec{\nabla}\cdot\vec{F}=0,
\end{equation}
where
$$\vec{F} = \frac{\gamma p}{\gamma - 1}\vec{v}+\frac{\vec{E}\times\vec{B}}{\mu}+\frac{1}{2}\rho v^2\vec{v}.$$
Taking the integral over the whole domain and making use of Gauss' divergence theorem, Equation \eqref{eq:total_energy_equation} can be written as
\begin{equation}
    \label{eq:total_energy_equation2}
    \frac{dE_{tot}}{dt}=-\oint_{\partial S}\left(\frac{\vec{E}\times\vec{B}}{\mu}\right)\cdot\vec{\hat{n}}dl,
\end{equation}
where $E_{tot}$ gives the total energy in the domain $S$, $E_{tot}$ is defined as
\begin{equation}
    \label{eq:total_energy_equation0}
    E_{tot}=\int_S\frac{p}{\gamma - 1}+\frac{B^2}{2\mu}+\frac{1}{2}\rho v^2dS.
\end{equation}
Most of the terms in $\vec{F}$ can be eliminated because $\vec{v}\cdot\vec{\hat{n}}=0$ on every boundary. The Poynting flux term can be simplified further by making use of Ohm's law as follows:
\begin{equation}
    \label{eq:ohms_law}
    \vec{E}=\vec{j}/\sigma - \vec{v}\times\vec{B}.
\end{equation}
Therefore,
$$\begin{aligned}
\frac{\vec{E}\times\vec{B}}{\mu}\cdot\vec{\hat{n}}&=\eta\vec{j}\times\vec{B}\cdot\vec{\hat{n}}-\frac{1}{\mu}(\vec{v}\times\vec{B})\times\vec{B}\cdot\vec{\hat{n}}, \\
&=\eta\vec{j}\times\vec{B}\cdot\vec{\hat{n}}-\frac{1}{\mu}(\vec{v}\cdot\vec{B})\vec{B}\cdot\vec{\hat{n}}.
\end{aligned}$$
This term can be simplified further because $\vec{j}\times\vec{B}\cdot\vec{\hat{n}}=0$ on the boundary. To demonstrate this, we first show that $B_y$ does not change on the bottom boundary. Consider the $y$-component of the induction equation as follows:
$$\begin{aligned}
\frac{\partial B_y}{\partial t}&=\nabla\times(\vec{v}\times\vec{B})\cdot\vec{\hat{y}}+\eta\nabla^2B_y, \\
&=\frac{\partial}{\partial x}\left(v_xB_y-B_xv_y\right)+\eta\nabla^2B_y, \\
&=0,
\end{aligned}$$
where $x$-derivative term equals zero because $\vec{v}=0$ on the boundary so it is constant along the bottom boundary; the Laplacian term equals zero because initially $B_y=-B_0y/L_0$ and therefore remains zero for all time. Now consider the $y$-component of the Lorentz force on the bottom boundary as follows:
$$\begin{aligned}
\vec{j}\times\vec{B}\cdot\vec{\hat{y}}&= \left(\frac{1}{\mu}(\vec{B}\cdot\vec{\nabla})\vec{B}-\vec{\nabla}\left(\frac{B^2}{2\mu}\right)\right)\cdot\vec{\hat{y}},\\
&=\frac{1}{\mu}\left(B_x\frac{\partial}{\partial x}+B_y\frac{\partial}{\partial y}\right)B_y-\frac{\partial}{\partial y}\left(\frac{B^2}{2\mu}\right),\\
&=0,
\end{aligned}$$
where the $y$-derivatives are zero because $\vec{\hat{n}}\cdot\nabla=0$ on the boundary and the $x$-derivatives are zero because $B_y=B_y(y)$. Hence, Equation \eqref{eq:total_energy_equation2} can be written as\begin{equation}
    \label{eq:total_energy_equation3}
    \frac{dE_{tot}}{dt}=-\frac{1}{\mu}\int_{y=y_{min}}v_zB_zB_ydx,
\end{equation}
where the integral is taken only over the bottom boundary as this is where the driver is located. Since $B_y$ does not depend on $x$ it can be taken out of the integral to give\begin{equation}
    \label{eq:total_energy_equation4}
    \frac{dE_{tot}}{dt}=-\frac{B_y}{\mu}\int_{y=y_{min}}v_zB_zdx.
\end{equation}

\section{Total Alfv\'en wave energy evolution}
\label{sec:zenergy}
Using Equations \eqref{eq:continuity} - \eqref{eq:iduction} as well as Faraday's law it can be shown that the rate of change of Alfv\'en wave energy density is given by
\begin{equation}
    \begin{aligned}
    \label{eq:zenergy_density_evolution}
    \frac{\partial e_A}{\partial t} + \vec{\nabla}\cdot\vec{F}_A =\vec{v}\cdot\vec{\nabla}\left(\frac{B_z^2}{2\mu}\right) -\frac{1}{\sigma}\left(\frac{\vec{\nabla} B_z}{\mu}\right)^2,
    \end{aligned}
\end{equation}
where $\vec{F}_A$ is the Alfv\'en wave energy flux and is given by
\begin{equation}
    \vec{F}_A=\frac{1}{2}\rho v_z^2\vec{v}+\vec{E}\times\vec{B}_z.
\end{equation}
In Appendix \ref{sec:total_energy_evolution} it was shown that $\vec{j}\times\vec{B}\cdot\vec{\hat{n}}=0$ on the boundary, hence if $v_x=v_y=0$ on the boundary then it can be shown that
$$\vec{E}\times\vec{B}_z\cdot\vec{\hat{n}}=\vec{E}\times\vec{B}\cdot\vec{\hat{n}},$$
on the boundaries of the domain. Consequently, by taking the integral of Equation \eqref{eq:zenergy_density_evolution} and substituting Equation \eqref{eq:total_energy_equation4} the following equation is obtained:
\begin{equation}
    \label{eq:zenergy_evolution}
    \frac{dE_A}{dt}=\frac{dE_{tot}}{dt}+\int_S\vec{v}\cdot\vec{\nabla}\left(\frac{B_z^2}{2\mu}\right)-\frac{1}{\sigma}\left(\frac{\vec{\nabla}B_z}{\mu}\right)^2dS.
\end{equation}

\section{Harmonic periods}
\label{sec:harmonic_time_periods}

The goal of this appendix is to derive an expression for the harmonic series associated with each field line as a function of the vector potential ($A$) at the initial time step. To do this, a change of coordinates is used to rewrite the wave equation in a form such that the wave speed is constant. For now, an expression is derived for the first quadrant where $A\ge0$. Symmetry arguments can be used to derive the formula for the other quadrants. The change of coordinates is given by
$$x=\sqrt{\hat{A}}L_0e^s,\ y = \sqrt{\hat{A}}L_0e^{-s},$$
where
\begin{equation}
    \label{eq:vector_potential}
    A=A_0\hat{A}=\frac{B_0}{L_0}xy\end{equation}
and $A_0=B_0L_0$. The linearised ideal wave equation is given by
\begin{equation}
    \label{eq:wave_equation}
    \frac{\partial^2v_z}{\partial t^2}=\frac{(\vec{B}_0\cdot\vec{\nabla})^2}{\mu\rho_0}v_z,
\end{equation}
Using the fact that
$$\begin{aligned}
\vec{B}_0\cdot\vec{\nabla}&=\frac{B_0}{L_0}\left(\frac{dx}{ds}\frac{\partial}{\partial x}+\frac{dy}{ds}\frac{\partial}{\partial y}\right)=\frac{B_0}{L_0}\frac{\partial}{\partial s},
\end{aligned}$$
the wave equation (Equation \eqref{eq:wave_equation}) can be rewritten as
\begin{equation}
    \label{eq:wave_equation3}
    \frac{\partial^2v_z}{\partial t^2}=\frac{B_0^2}{\mu\rho_0L_0^2}\frac{\partial^2 v_z}{\partial s^2}.
\end{equation}
Thus, the harmonic periods are given by
\begin{equation}
    \tau_n = \frac{2}{n}\frac{L_0}{v_{A0}}(s_2-s_1),
\end{equation}
where $s_1$ and $s_2$ are the values of $s$ at each of the footpoints. In the first quadrant, at $s_1$, $y=y_{max}$ and at $s_2$, $x=x_{max}$ , therefore
$$s_1=-\log\left(\frac{y_{max}}{L_0\sqrt{\hat{A}}}\right),\  s_2=\log\left(\frac{x_{max}}{L_0\sqrt{\hat{A}}}\right).$$
Finally, the harmonic periods are given by
\begin{equation}
    \tau_n = \frac{2}{n}\frac{L_0}{v_{A0}}\log\left(\frac{A_0}{L_0^2}\frac{x_{max}y_{max}}{A}\right),
\end{equation}
using symmetry arguments it can be shown that the formula for all quadrants is given by
\begin{equation}
    \label{eq:harmonic_time_periods}
    \tau_n = \frac{2}{n}\frac{L_0}{v_{A0}}\log\left(\frac{A_0}{L_0^2}\frac{x_{max}y_{max}}{|A|}\right).
\end{equation}

\section{Resonance locations}
\label{sec:resonance_locations}

The driving period is given by
\begin{equation}
    \label{eq:driving_time}
    \tau_{driv}=\frac{L_0\sqrt{\mu\rho_0}}{B_0}4\log\left(2\right).
\end{equation}
The driving period equals one of the harmonic periods, where $\tau_n=\tau_{driv}$ and
$$\frac{|A|}{A_0}=\frac{|xy|}{L_0^2}=\frac{x_{max}y_{max}}{L_0^24^n},$$
\begin{equation}
    \label{eq:resonance_locations}
    \implies \frac{xy}{L_0^2}=\pm4^{1-n},\quad y\le0,
\end{equation}
where $y\le0$ because the driver is imposed on the bottom boundary.

\end{appendix}

\bibpunct{(}{)}{;}{a}{}{,} 
\bibliographystyle{aa}        
\bibliography{bibliography.bib}           

\end{document}